\documentclass{aa}
\usepackage{epsfig,amsmath,amssymb}

\def\calN{{\cal N}}
\def\msol{M_\odot}

\def\mjup{M_{\rm J}}
\def\rjup{R_{\rm J}}
\def\mearth{\,{\rm M}_\oplus}

\def\mcore{\,{\rm M}_{\rm core}}
\def\yeff{\,{\rm Y}_{\rm equiv}}
\def\zenv{\,{\rm Z}_{\rm env}}

\def\simgr{\,\hbox{\hbox{$ > $}\kern -0.8em \lower 1.0ex\hbox{$\sim$}}\,}
\def\simle{\,\hbox{\hbox{$ < $}\kern -0.8em \lower 1.0ex\hbox{$\sim$}}\,}
\def\beq{\begin{equation}}
\def\eeq{\end{equation}}

\begin{document}

\title{Structure and evolution of super-Earth to super-Jupiter exoplanets:
I. heavy element enrichment in the interior}

 \author{I. Baraffe\inst{1}, G. Chabrier\inst{1}, T. Barman\inst{2}
}

\offprints{I. Baraffe}

\institute{ \'{E}cole normale sup\'erieure de Lyon, CRAL (CNRS), 46 all\'ee d'Italie, 69007 Lyon,\\ Universit\'e de Lyon, France (ibaraffe, chabrier@ens-lyon.fr)
\and
Lowell observatory, 1400 West Mars Hill Road, Flagstaff, AZ 86001, USA (barman@lowell.edu)
}

\date{Received /Accepted}

\titlerunning{From super Earth to super Jupiter exoplanets.}
\authorrunning{Baraffe, Chabrier \& Barman}
\abstract {}
{We examine the uncertainties in current planetary models and we quantify
their impact on the planet cooling histories and mass-radius relationships. These uncertainties include
(i) the differences between the various equations of state used to characterize the heavy material  thermodynamical properties, 
(ii) the distribution of heavy elements within planetary interiors, (iii) their chemical composition and (iv) their thermal contribution to the planet evolution. Our models, which include a gaseous H/He envelope, are compared with models of solid, gasless Earth-like planets in order to examine the impact of a gaseous envelope on the cooling and the resulting radius.} 
{We find that for a fraction of heavy material larger than 20\% of the planet mass, the distribution of the heavy elements
in the planet's interior affects substantially the evolution and thus the radius at a given age.
For planets with large core mass fractions ($\simgr$ 50\%), such as the Neptune-mass transiting planet GJ436b, the contribution
of the gravitational and thermal energy from the core to the planet cooling
history is not negligible, yielding a $\sim$ 10\% effect on the radius after 1 Gyr.}
{We show that the present mass and radius determinations of the massive planet Hat-P-2b require at least
200 $\mearth$ of heavy material in the interior, at the edge of what is currently predicted by the core-accretion model for planet formation. As an alternative avenue for massive planet formation,
we suggest that this planet, and similarly HD 17156b, may have formed from
collisions between one or several other massive planets. This would explain these planet unusual
high density and high eccentricity.
We show that if planets as massive as $\sim$ 25 $\mjup$ can form, as predicted by improved core-accretion
models, deuterium is able to burn in the H/He layers above the core, even for core masses as large
as $\sim$ 100 $\mearth$. Such a result highlights the confusion provided by a definition of a planet
based on the deuterium-burning limit. 
We provide extensive grids of planetary evolution models from 10 $\mearth$ to
 10 M$_{\rm Jup}$, with 
various fractions of heavy elements. These models provide a reference to analyse the transit discoveries expected from the CoRoT and Kepler missions and to infer the internal composition of these objects.}


\maketitle

\section{Introduction}

The number of newly discovered exoplanets transiting their parent star keeps increasing continuously, revealing a remarkable diversity in mean densities for planet masses ranging from Neptune masses to
several Jupiter masses. A large fraction of these transits exhibit a mean density significantly larger than 
that of an object essentially composed  of gaseous H/He, like brown dwarfs or stars, indicating a composition substantially enriched in heavy elements. The first compelling evidence for such a significant enrichment was provided by the discovery of a Saturn mass planet, HD149026b, with such 
a small radius that 2/3 of the planet's mass must be composed of elements heavier than He 
(Sato et al. 2005).
Another remarkable discovery is the case of GJ 436b, a $\sim$ 22 $\mearth$ Neptune-like planet with
a radius comparable to that of Uranus or Neptune (Gillon et al. 2007a). Such a radius implies an inner structure composed by more than 90\% of heavy elements. That exoplanets can be substantially enriched in heavy material such as rock or ice\footnote{Under usual planet formation conditions, the word "rocks" refers primarily to silicates (Mg-, Si- and O-rich compounds) whereas the term "ice"
includes collectively H$_2$O, CH$_4$ and NH$_3$, water being the most important of these three
components. As will be discussed in \S3, the term "ice" may be inappropriate in some cases, as water
could be under a liquid or gaseous form.} is not a surprise, since this is a well known property of our own Solar System planets. 
Moreover, the presence of an icy/rocky core and of oversolar metallicity in the envelope is an expected consequence of the most widely accepted planet formation scenario, the so-called core-accretion model (Alibert et al. 2005a, and references therein). 

Given this expectation, it becomes mandatory
to take into account heavy element enrichment in planetary models devoted to the analysis and the
identification of
current and forthcoming observations of extra-solar planets. Many efforts are now devoted to the
modeling of massive terrestrial planets, essentially composed of solid material 
(Valencia et al. 2006; Sotin et al. 2007; Seager et al. 2007) and jovian planets with H/He envelope and a substantial  metal enrichment (Baraffe et al. 2006; Guillot et al. 2006; Burrows et al. 2007; Fortney et al. 2007).
Because of remaining uncertainties in the input physics describing the planetary structures, and of many unknown quantities such as the total amount of heavy elements, their chemical composition and their distribution within the planet's interior, these models are based on a number of assumptions and thus retain some, so far unquantified, uncertainties. 
The main goal of the present paper is to analyse and quantify these uncertainties  and to explore as precisely as possible the impact of the heavy material contribution on the planet structure and evolution. We will focus on planets with mass larger than 10 $\mearth$, the expected limit for the
 gravitational capture of a gaseous H/He envelope and atmosphere (Mizuno 1980; Stevenson 1982; Rafikov 2006; Alibert et al. 2006), which has a significant impact on the evolution. Below this limit mass, the objects are essentially solid bodies with no or a teneous gaseous envelope and their mass-radius relationship has been studied recently by several authors (Valencia et al. 2006, 2007; Sotin et al. 2007; Seager et al. 2007).
This study is motivated by the level of accuracy on planetary mass and radius
measurements which is now reached with ground-based (HARPS, VLT) and  space-based instruments (HST, SPITZER). Observations  are expected to reach an unprecedented level of precision in the near-future with COROT, KEPLER, and on a longer term with GAIA. The latter project will measure distances and thus stellar radii with high precision, removing one of the main sources of uncertainty in planetary radius measurements. This race
for precision is motivated by the possibility to infer with the best possible accuracy
the inner composition of an exoplanet, with
the aim to better understand planet formation and to identify the presence of astrobiologically
important material such as liquid water. In this context, it is crucial to quantify the uncertainties in the structure
and evolution planetary models used to analyse these observations.
In section \S2 and \S3, we examine the main input physics and assumptions used in structure and evolutionary models available in the literature. In section \S4, we analyse quantitatively the impact of  these assumptions. We focus  on specific cases such as HD149026b and GJ436b in \S 5 and on
super Jupiter planets in \S 6. Our various planetary models, covering a wide mass range and including different levels of heavy element enrichments, are presented in \S 7. Discussion and perspectives follow in \S 8. 

\section{Uncertainties and assumptions in the modelling of extra-solar planets}

\subsection{Distribution of heavy material within the planet}

According to a recent study devoted to the structure of our giant planets (Saumon \& Guillot 2004), Jupiter should have a total amount of heavy elements ranging from
8 to 39 $\mearth$, {\it i.e} a metal mass fraction $Z \sim $  2.5\% to 12\%,
with a maximum core mass of 11 $\mearth$  and a maximum envelope metal mass 
fraction $Z_{\rm env}$ = 12\%.
For Saturn,  the same study suggests a total mass of heavy elements
 ranging from 13 to 28 $\mearth$,
 {\it i.e} $Z \sim $  13\% to 29\%, with a maximum core mass of 22 $\mearth$ and a maximum
$Z_{\rm env} \sim $  8\%. These properties 
can be understood within the standard general framework of giant planet formation : as a core is growing in mass due to accretion of planetesimals, it reaches a critical mass around $\sim 6$-10 $\mearth$ above which gas accretion begins (Mizuno 1980; Stevenson 1982; Rafikov 2006). During this gas accretion phase, planetesimals are still accreted and are either destroyed in the gas envelope or are falling onto the core, leading to further increase of the core mass. The fate of these accreted planetesimals, disrupted in the
envelope or accreted onto the dense core, depends 
on their size, an unknown parameter in current planet formation models, and on the envelope mass. This general picture thus predicts that planets should have
a dense core of heavy material (rock, water/ice) of, at least, a few Earth masses and can show different levels of heavy element enrichment in their envelope, depending on the accretion history (amount of gas accreted, size of planetesimals, etc...).
Current models based on the core-accretion scenario are able to match the core mass and the envelope metal enrichment derived for Jupiter and Saturn (Alibert et al. 2005b), but predict in some cases
much larger heavy element enrichments in the envelope, depending on the planet's mass, with
values as large as $Z_{\rm env}\simgr 50\%$ for Neptune-size planets ($\sim 10$-20 $\mearth$)
(Baraffe et al. 2006). 

Despite this widely accepted picture, current extra-solar planet models often simply assume
that all heavy elements are located in the central core and that the envelope is either metal-free, $Z_{\rm env}=0$, (Burrows et al. 2007; Fortney et al. 2007), or has a solar metallicity\footnote{Note also that different envelope
helium mass fractions have been assumed, with $Y=0.25$ (Burrows et al. 2007) or $Y$=0.28 (Fortney et al. 2007).}, $Z_{\rm env}=Z_\odot$ 
(Guillot et al. 2006). This simplification is based on the assumption that whether the heavy elements are located in the core or in the envelope should not affect the planet's evolution. The validity of such an assumption, however, has
 never been examined, as will be done in the present paper. 


\subsection{Thermal contribution to the EOS}

An other simplification found in planet modelling is the 
use of temperature-independent EOS, assuming that the heavy element material is at zero-temperature
or at a uniform, low temperature
(Seager et al. 2007; Fortney et al. 2007). 
This is certainly a good assumption when examining the structure of terrestrial-like planets,
composed essentially of solid (rocky/icy) material (Valencia et al. 2006; Sotin et al. 2007). This assumption, however, is not necessarily valid for
the evolution of more massive planets. For these objects,
 the thermal and gravitational energy contributions of the core to the planet's cooling history are usually ignored
 and it is important to examine the impact of such a simplification.
 

\subsection{ Heat transport} 

Finally, a conventional assumption in planet modelling is to assume that the interiors (at least the gaseous envelope) of giant planets are homogeneously mixed, due to the dominant and supposedly
efficient transport mechanism provided by large-scale convection. Accordingly, the internal temperature gradient is given by the adiabatic gradient, since superadiabaticity is negligible.  The validity of this
 assumption, however, has been debated for the interior of our own jovian planets for already some time
 (Stevenson 1985) and has been questioned again more recently in the context of transiting extra-solar planets (Chabrier \& Baraffe 2007). It should be kept in mind, however,
  that, even though the planetary internal heat
 flux can only be carried out by convection (Hubbard 1968), the assumption of large-scale, fully adiabatic
 convection in planetary interiors has never been proven to be correct and even slightly inefficient
 convection can have a major impact on the planet's structure and evolution (Chabrier \& Baraffe 2007).

\section{Equations of state for heavy elements} 

\subsection{The case of water} 

As quickly mentioned in the introduction, the generic term "ice" may often be inappropriate to describe
the thermodynamic state of water under planetary conditions. Indeed,
depending on the temperatures prevailing
in the parent protoplanetary nebula at the initial location of the planet embryo, water
could possibly be initially under the form of solid ice, but it could also be under
the form of  liquid or vapor. If initially solid, it may also melt or vaporize under the conditions prevailing in the planet's interior, or may also dissociate under the form of ionic melts at high pressures and temperatures (Schwegler et al. 2001). 
In fact, the phase diagram of heavy elements under the pressure and temperature conditions characteristic of the considered planet interiors is largely unknown, as only part of this diagram
is presently accessible to high-pressure experiments or
computer numerical simulations.
Water, for instance, the dominant component after H and He, is known to exhibit a complex
phase diagram with many stable or metastable (amorphous) phases and several triple and critical
points, because of the high flexibility of the hydrogen bonding. The melting line
of water at high pressure and high temperature has been probed experimentally up to $P=3.5\times 10^{11}$ dyn cm$^{-2}$ (0.35 Mbar) and $T=1040$ K. H$_2$O has been found to remain solid
at larger pressures and temperatures, suggesting that the melting line increases (in $T(P)$)
at higher temperatures and extends
up to at least $T\simgr 1600 $ K (Lin et al. 2005). Depending on its extrapolation at higher temperatures,
this line may intersect the internal profiles of some of our
planet models, at least in the Super-Earth and Neptune mass range, so that both liquid and solid H$_2$O might be present at some depth in the interiors of these objects, as predicted for ocean-planets (Selsis et al. 2007), while supercritical H$_2$O is more likely to be
liquid or gaseous in the hotter interiors of Saturn-like or larger planets.
As mentioned above, water is also found from shock-wave experiments and first-principles calculations to dissociate into $H_3O^++OH^-$ ion pairs above $\sim 2000$ K at 0.3 Mbar (Schwegler et al. 2001).
In fact, the distinction between "gas", "ice" and "rock"
may become meaningless under extreme conditions. In that case, what matters is the global
amount of heavy elements. The term "ice", or "water", may thus more generically refer to the
volatile forms of O, C and N, while "rock" refers primarily to silicates (Mg-, Si- and O-rich compounds)
and "iron" refers to the rest (metal, oxide, sulfide or substituting for Mg in the silicates) (Stevenson 1985).


\subsection{EOS available in the literature}

\begin{figure*}
\psfig{file=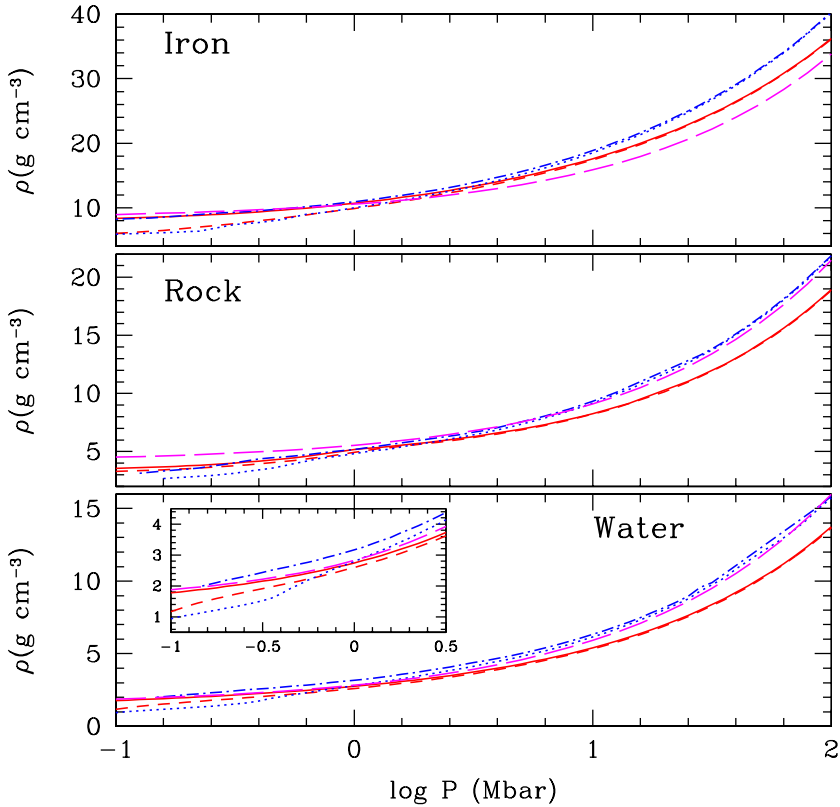,height=120mm,width=120mm} 
\caption{EOS for water, rock and iron  for two temperatures: T=300K for ANEOS (red solid line) and SESAME (blue dash-dotted curve); T=6000K for ANEOS (red short-dashed curve) and SESAME (blue dotted curve). The long-dashed (magenta) curves correspond to the zero-temperature EOS of Seager et al. (2007; their Table 3). The inset in the lower panel (water) shows a zoom of the $P(\rho)$ relation for $\log \, P \, <  \, 0.5$}
\label{fig1}
\end{figure*}

For Earth-like planets, models can test a variety of complex heavy element compositions inspired by the knowledge of the structure of our own Earth (Valencia et al. 2006, 2007; Sotin et al. 2007). 
Moreover, the EOS of  materials which may be found in planetary interiors (water, iron, dunite or olivine, etc...) are reasonably well determined at zero-temperature or at 300 K (see Seager et al. 2007). Unfortunately,
for larger planets, the exploration of internal compositions is
restricted to a few materials for which EOS are available and cover a large enough range of pressures and temperatures. The two most widely used EOS in this context (Saumon \& Guillot 2004; Baraffe et al. 2006; Fortney et al. 2007; Burrows et al. 2007) are ANEOS (Thompson \& Lauson 1972) and SESAME (Lyon \& Johnson 1992), which describe the thermodynamical properties of {\it water}, {\it rocks} (olivine or dunite, {\it i.e} Mg$_2$SiO$_4$, in ANEOS; a mixture of silicates and other heavy elements called "drysand" in SESAME) and {\it iron}.
Figure \ref{fig1} shows a comparison between these two EOS for the three aforementioned materials for two temperatures, namely $T=300$ K and $T=6000$ K. Comparison is also shown with the zero-temperature EOS presented in Seager et al. (2007) for water, perovskite (MgSiO$_3$) and iron. These authors use fits to experimental data at low pressure ($P \simle$ 2 Mbar) 
and an improved Thomas-Fermi-Dirac theory at high pressure ($P \simgr 100$ Mbar), where the contribution of the degenerate electron fluid becomes dominant. As stressed by these authors, the main difficulty is to bridge the gap in the pressure regime $2 \, {\rm Mbar}  \simle P \simle$ 100 Mbar ($2 \times 10^{12}$ - 10$^{14}$ dyne cm$^{-2}$), which is the most relevant for planetary interiors. Their model EOS represents some improvement upon ANEOS and SESAME, in particular for water, but is valid
only at zero-temperature.

The presently used EOS models (ANEOS and SESAME), and
thus the inferred planet internal structures, thus retain a significant degree of uncertainty in the experimentally unexplored high-P and high-T domains.
In these regimes, both EOS models, as well as the one used by Seager et al. (2007), are based on
interpolations
between experimental data at low or moderate density/temperature and well-known asymptotic limits, in general
the Thomas-Fermi or more accurate density-functional type models, in the very high density, fully ionized limit. One can only hope that these
interpolations do not depart too much from reality, as might be the case, for instance, if first-order phase transitions, which imply density and entropy discontinuities, occur in the regions of interest. For the static properties, this assumption is
probably reasonable at high temperatures, but might be more questionable near
melting lines. When addressing the transport properties, like e.g. the thermal diffusivity or kinematic viscosity, the results in the interpolated regime are definitely more doubtful. 


As shown in Figure \ref{fig1}, at low (room) temperature, the various EOS agree reasonably well.
For water, the
agreement lies within less than 5\%
at $P$=0.1 Mbar and within less than 16\% at $P=100$ Mbar. 
For "rocks", keeping in mind that this term refers to different compositions in the various EOS,  the
agreement is comparable, although the difference can reach 27\% at P=0.1 Mbar. For iron,
the three EOS agree well at low pressure but differences as large as $\sim$ 20\% can occur at P=100 Mbar between SESAME and Seager et al. (2007). Such cool temperature conditions, however, are more relevant
to Earth-like planets than to the ones explored in the present study.

Figure \ref{fig1} also shows, for  ANEOS and SESAME, the variation of $P(\rho)$ with temperature,
for conditions more suitable to our planetary interior conditions. The
thermal contributions predicted by these EOS are significant at $P \lesssim 1$ Mbar, with
a $\sim$ 40-60\% difference in $P(\rho)$ between the $T=300$ K and $T=6000$ K isotherms,
for iron and water, respectively. The differences keep increasing significantly for $T > 6000$ K. 
It is instructive to quantify the impact of these thermal contributions on the evolution of our planets, and to determine
whether neglecting the temperature dependence of the EOS is consequential or not. 
 
\begin{figure}
\psfig{file=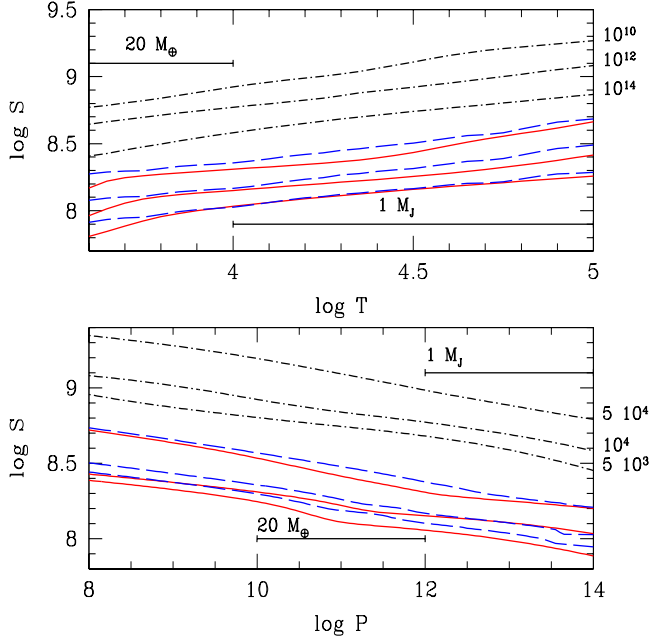,height=100mm,width=88mm} 
\caption{Pressure (in dyne cm$^{-2}$) and temperature (in K) dependence of the entropy (in erg/g/s), as obtained for water for the ANEOS EOS
(solid line) and the SESAME EOS (dash-line). The entropy for a H/He fluid, with $Y=0.275$ obtained with
the SCVH EOS is shown for comparison (dash-dotted line). Upper panel: for each EOS, three isobars
are displayed (the corresponding pressure is given on the right hand side of the figure). Lower panel: three isotherms
are shown for each EOS (corresponding temperatures given on the right hand side of the figure). 
The $T$ and $P$ domains characteristic of Neptune-like and Jupiter-like planet interiors are indicated.}
\label{figPTS}
\end{figure}

Whereas $P(\rho)$ is the relevant quantity for the {\it structure}, the relevant one for the {\it evolution} is
the entropy. Figure \ref{figPTS} portrays the P- and T-dependence of the
entropy, for three isotherms and isobars, for water, for the ANEOS and SESAME EOS, under
conditions relevant to the planets of present interest. The values for
the H/He fluid (EOS of Saumon et al. 1995; hereafter SCVH EOS) are also displayed for comparison. We see that, whereas the EOS agree
reasonably well under Jovian-planet conditions, as expected as they reach the asymptotic, high-P,
high-T
regime accurately described by Thomas-Fermi-Dirac or more accurate density-functional theories,
the difference can be substantial for conditions characteristic of the Neptune-mass domain.  Indeed, for this latter case, most of the interior lies in the interpolated regime where guidance from either experiments or
numerical simulations is lacking. These differences between the EOS, of course, are amplified for the quantities involving the
derivatives of the entropy, such as the adiabatic gradient or the specific heat.

\subsection{Treatment of  metal enrichment in the core and in the envelope} 
 
Our goal is to examine the impact on the planetary models and the inferred mass-radius relationships due to uncertainties in 
(i) the distribution of heavy elements  within the planet interiors, (ii) their dominant chemical composition, (iii) the
different EOS describing their thermodynamical properties and (iv) their thermal contribution.  To achieve this goal, we have implemented the ANEOS and SESAME EOS for water, rock
and iron (or what is so-denominated) in our planetary evolution code. Both EOS provide all thermodynamic quantities relevant to the evolution of planets, including internal energy,  entropy (in ANEOS) or free energy (in SESAME), and all relevant derivatives. ANEOS also provides Rosseland and conductive opacities. When the SESAME EOS is used,
conductive opacities are calculated according to Potekhin (1999). The evolutionary calculations, including the
presence of a dense core, proceed as described
in Baraffe et al. (2006): the structure equations
are integrated from the center to the surface; at the core boundary, the EOS is switched to the
one characteristic of the gaseous envelope, and the change in chemical composition yields a density jump, but continuity in pressure and temperature is enforced.

To account for the thermodynamic contributions of heavy elements in the H/He envelope, we have tested two
different procedures.(1) Following the method described in Chabrier et al. (1992) and used in
 Baraffe et al. (2006), we  first mimic the presence of metals with mass fraction $Z$ by an equivalent helium fraction $Y_{\rm equiv} = Y+Z$ in the H/He SCVH EOS, with $Y$ the real helium mass fraction in the envelope. As mentioned in
 Chabrier et al. (1992), this approximation is reasonable as long as the mass fractions $Y$ and
 $Y_{\rm equiv}$ are small compared with unity and both $\rho_Z$ and $\rho_{He}$ are large compared
 with $\rho_H$, where $\rho_i$ denotes the mass density of the i-component at pressure $P$.
(2) The second, more general approach to describe the EOS of a mixture of various species in the absence of a reliable theory is to apply exactly the additive volume law (hereafter AVL), which is
exact in the ideal gas limit, without restriction on the species mass fractions and densities (see Fontaine et al. 1977 and Saumon et al. 1995 for extensive discussions of the validity of the AVL). In this approach, the
interactions between the three different fluids, namely hydrogen, helium and the heavy element
component, are
not taken into account (but interactions between particles in {\it each} of these fluids are treated properly)
and the EOS of the mixture is simply the mass-weighted interpolation of each species contribution at
constant intensive variables, $P$ and $T$, plus the ideal entropy of mixing for the entropy term.
Within the ideal volume law, the mass density of the mixture of a H/He fluid with a helium mass fraction $Y$ plus some heavy element material with mass fraction $Z$ at pressure $P$ and temperature $T$
thus reads:

\beq
{1 \over \rho(P,T)} = {(1-Z) \over \rho^{H/He}(P,T)} + {Z \over \rho^Z(P,T)}
\eeq
The extensive variables, e.g. the internal energy and specific entropy, thus read:

\beq
U(P,T) = (1-Z)\,U^{H/He}(P,T) + Z\,U^Z(P,T)
\eeq
\beq
S(P,T) = (1-Z)\,S^{H/He}(P,T) + Z\,S^Z(P,T) + S_{\rm mix}(P,T)
\eeq
where the EOS of the $H/He$ component is given by the SCVH EOS while
the one of the $Z$-component is described by either the ANEOS or the SESAME EOS. 
All along the present calculations, we have taken the cosmic helium fraction $Y$=0.275.
The details of the calculation of the contribution due to the ideal entropy of mixing are given in Appendix A. 
This term is found to contribute non negligibly to the total entropy $S$. Depending on the mixture and the $P$-$T$ range, it can amount to 10\%-20\% of $S$. This is consequential when calculating the
{\it internal structure} of a planet, whose interior is essentially isentropic, at a given age.

For the evolution, the relevant quantities  are the
variation of the entropy with time and its derivatives w.r.t $P$ and $T$, which give the adiabatic gradient:

\beq
\nabla_{\rm ad} = ({\partial \log S \over \partial \log P})_{_T} / ({\partial \log S \over \partial \log T})_{_P}
\eeq

Given the fact that the internal composition of the planet ($N_H, N_{He}, N_Z$, where
$N_i$ denotes the number of particles of species $i$), remains constant along the evolution,
the only $P$- and $T$-dependence of the mixing entropy term arises from the variation of the degree
of ionization and thus of the abundance of free electrons. The degree of ionisation of the heavy element component, however, is unknown, and its
variation with temperature has been ignored in the present
calculations. Therefore, only the variation with temperature of the number of electrons provided by H and He contribute to the variation with time of the mixing entropy. Within this limitation, the variation of
$S_{\rm mix}$ with $P$ and $T$ is found to be negligible for planets in the mass range
of interest, for all levels of 
metal enrichment in the envelope, so that the entropy of mixing term does not contribute significantly to the
{\it evolution}.




\section{ Effect of the different treatments and distributions of heavy elements on the planet cooling history}

In this section, we analyse the impact of the localisation of the heavy elements within the planet
on its structure and evolution. The heavy elements are distributed either in the core
or in the gaseous envelope, and we consider  in the present section mass fractions with $Z \le 50\%$. This
encompasses the level of enrichment of  our giant planets 
Jupiter and Saturn and of  previous theoretical studies
devoted to the analysis of transit planets
(Guillot et al. 2006; Burrows et al. 2007). 
The effect of  larger fractions of heavy elements ($Z> 50\%$) will be explored in the next section, with the analysis of specific cases of planetary transits.
To cover the largest possible range of conditions, we explore these effects on a Neptune mass (20 $\mearth$) and  a Jovian mass planet (1 $\mjup$ = 318 $\mearth$). As a test case, we restrict the analysis in this section to planets irradiated by a Sun at 0.045 AU, since the observational determination of the radius is
presently accessible only to transiting, short-period planets. We use outer boundary conditions derived from a grid of irradiated atmosphere models with solar metallicity (Barman et al. 2001).  These models take into account the incident stellar flux, defined by $F_{\rm inc}= {f \over 4} ({R_\star \over a})^2 F_\star$, with $R_\star$ and $F_\star$ the radius and the flux of the parent star,  respectively, and $f$ the redistribution factor. 
The Barman et al. (2001) models use $f$=2, corresponding to the incident flux being redistributed only over the dayside of the planet. Recent observations of the day-night contrast of exoplanets with Spitzer (see e.g. Knutson et al. 2007)
and atmopsheric circulation models (see the discussion in Marley et al. 2007), however,
seem to favor a redistribution over the entire planet's surface, 
{\it i.e} $f$=1. For the specific case of HD209458b, Baraffe et al. (2003) 
found out that such a variation of $F_{\rm inc}$ by a factor 2 has
a significant effect on the outer atmospheric profile, but a 
small effect on the planet evolution.
In a forthcoming paper, we will explore in more details  the effects of the 
redistribution factor $f$,
of different levels of irradiation and of heavy element 
enrichment in the atmosphere.

\subsection{Neptune-mass planets}

\begin{figure}
\psfig{file=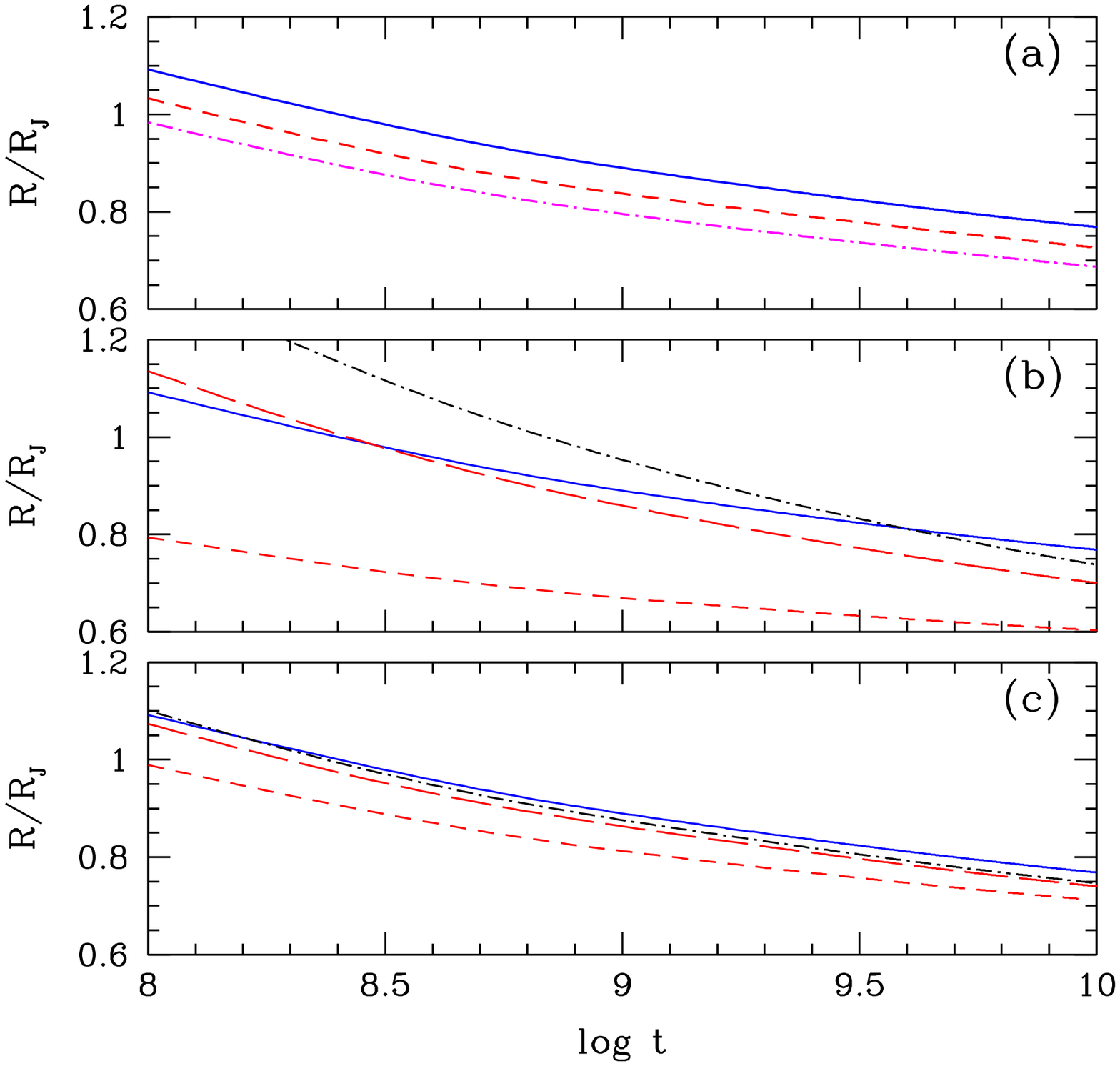,height=110mm,width=80mm} 
\caption{Effect of the composition and internal distribution of heavy elements on the radius
 evolution  for a planet of mass $M_P=20 \mearth$ with a total heavy element mass
 fraction $Z=(M_Z/M_P)$=50\%. 
{\bf (a):} Models with $\mcore = 10 \mearth$ of water (solid line), rock (dashed-line)
and iron (dotted line), and $\zenv$=0.
{\bf (b):} Models with no core and heavy elements distributed over the entire planet.
Dash-dotted line: mixture of H/He + $Z$ described by  the SCVH EOS with 
$\yeff$=0.275 + 0.5=0.775. Long-dashed line: mixture of H/He + water (SESAME EOS)
using the additive volume law. Short-dashed line: 
mixture of H/He + water (ANEOS EOS) using the additive volume law. For comparison,
the model with $\mcore = 10 \mearth$ of water and $\zenv$=0 is shown by the solid line.
{\bf (c):} More realistic models with a 8 $\mearth$ core of water (ANEOS EOS) and a $\zenv$ = 16.6 \% heavy element enrichment in the envelope. Dash-dotted line: mixture of H/He + $\zenv$ described by  the SCVH EOS with  $\yeff$= 0.275 + 0.166 = 0.441. Long-dashed line: mixture of H/He + water (SESAME EOS) using the additive volume law. Short-dashed line: 
mixture of H/He + water (ANEOS EOS) using the additive volume law.
Also displayed is the model with $\mcore = 10 \mearth$ of water and $\zenv$=0 (solid line).
}
\label{fig2}
\end{figure}

Figure \ref{fig2} shows the evolution of the radius with time for a 20 $\mearth$ planet with $Z$=50\%, for different heavy element materials and for different distributions of these latter in the planet interior. 
We first consider  models with a 10  $\mearth$ core and $\zenv$=0, i.e. {\it all} heavy elements are
located in the central core of the planet. The results for pure
 water (solid line), rock (dashed line) and iron (dash-dotted line) cores are displayed in
Fig. \ref{fig2}a, based on the ANEOS EOS for these materials. A comparison of these results with
those obtained with the SESAME EOS for  the same material (water and drysand) shows less
than a 1\% difference on R at a  given time. For iron, the SESAME EOS does not provide
the free energy and thus the entropy, the adiabatic gradient and other quantities required
for the evolution. However, according to the comparison between the ANEOS and SESAME EOS
portrayed in Fig. \ref{fig1},
we do not expect significant differences on the evolution between the different iron EOS.

For the aforementioned global heavy element enrichment, $Z=50\%$, the
nature of the core material affects the cooling, and thus the radius evolution at the $\sim$10\% level.
 We will see in \S 5,
however, that for larger heavy element mass fractions,
 the impact of the core composition can be more severe.
 At 1 Gyr, the radius of the planet with a pure rocky (iron) core is smaller by 6\% (11\%) compared to the pure water core case (see Table 1) . For water and iron, the ANEOS EOS indicates whether the material is in a solid, liquid or melt phase. For dunite, this information is not provided. In all the models, water is
 always found to be in a liquid state. However, as mentioned in \S3.1, most of our planet interiors lie in
 the extrapolated, supercritical region of the phase diagram; the precise state of water in this regime
depends on the extrapolation of the melting curve and thus is presently undetermined. Moreover, as
mentioned earlier, at high pressure and temperature, the meaning of "water" becomes loose and refers
more generally to protons and oxygen nuclei.
 The planet with an iron core undergoes a phase change from liquid
 to solid, according to ANEOS, at an age of $\sim$ 1 Gyr. This transition starts in the most central part of the core, at temperatures $T \sim 1.7\times 10^4$ K and $\rho \sim 30 $ g.cm$^{-3}$. 
  
  \begin{table*}
\caption{ Radius and central thermodynamic properties of a planet of mass $M_P=20 \mearth$ at 1 Gyr with a total heavy element mass fraction $Z=(M_Z/M_P)$=50\%. Results for different heavy element compositions (water=W, rock=R, iron=I) and EOS (aneos=a, sesame=s, SCVH=SC) and different heavy element distributions are given.}
\label{tab1}
\begin{tabular}{clllllll}
\hline
\hline 
 $\mcore$ & EOS (core) & $\zenv$ & EOS(env) & R$_{\rm p}$ & $T_{\rm c}$  & $\rho_{\rm c}$  \\
    ($\mearth$) &   & & env. & ($\rjup$) & (K) & (g cm$^{-3}$) \\
 \hline
 \hline
10 & W-a & 0 & SC & 0.890 &  1.1 10$^4$ &    4.11 \\
10 & W-s & 0 & SC & 0.901 &  1.5 10$^4$ &    4.85\\
10 & R-a & 0 & SC & 0.836 &  8.1 10$^3$ &    9.07\\
10 & R-s & 0 & SC & 0.838 &  1.8 10$^4$ &   10.64\\
10 & I-a & 0 & SC & 0.799 &  1.7 10$^4$ &   31.20\\
0 &  & 0.5 & SC (0.775$^a$) & 0.953 &  1.2 10$^4$ &    0.58\\
0 &    & 0.5 & SC+W-a & 0.669 &  7.6 10$^3$ &    1.03\\
0 &      & 0.5 & SC+W-s & 0.858 &  10$^4$ &    0.73\\
8 & W-a & 0.166 & SC (0.441$^a$) & 0.876 &  1.2 10$^4$ &    3.85\\
 8 & W-a & 0.166 & SC+W-a & 0.811 &  1.1 10$^4$ &    3.93 \\
 8 & W-a & 0.166 & SC+W-s & 0.862 &  1.1 10$^4$ &    3.87\\
 \hline
 $^a$ Value of $\yeff$.
 \end{tabular}
\end{table*}

\begin{figure}
\psfig{file=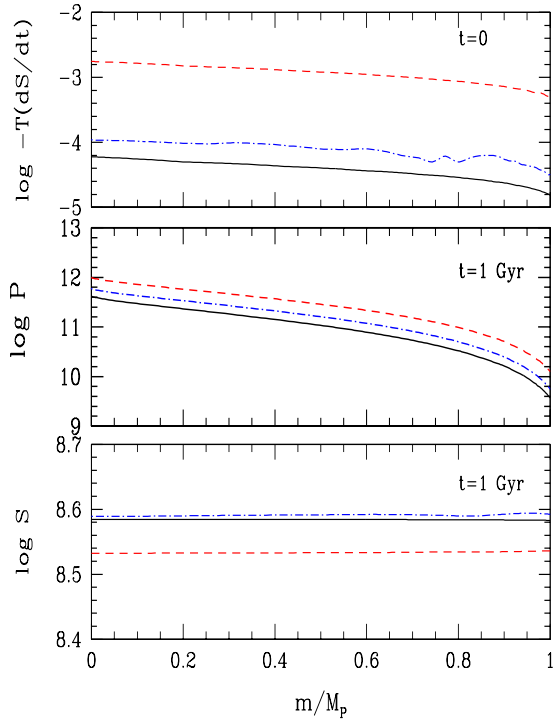,height=110mm,width=80mm} 
\caption{Effect of the EOS on the inner profile as a function of fractional mass for  a planet of 20 $\mearth$ with
a mass fraction $Z$=50\% of heavy elements distributed over the entire planet. In all panels,
the curves correspond to:  a mixture of H/He + $Z$ described by:  the SCVH EOS with 
$\yeff$=0.275 + 0.5=0.775 (solid line); a mixture of H/He + water (SESAME EOS)
using the additive volume law (dash-dotted line); a mixture of H/He + water (ANEOS EOS) using the additive volume law (dashed line).
{\it Upper panel:} Inner profile of the local gravothermal energy, $-TdS/dt$ (in erg/g/s), at the beginning of the evolution. 
{\it Middle panel}: Pressure profile (in dyne/cm$^2$) at 1 Gyr.
{\it Lower panel}: Specific entropy profile (in erg/g/K) at 1 Gyr.
}
\label{fig3}
\end{figure}

 Figure \ref{fig2}b illustrates the effect of the localisation of the heavy elements within the planet's interior,
 with two limiting assumptions: (A) all heavy elements are in the core ($\zenv$=0)
 and (B) they are distributed all over the planet's interior ($\mcore$=0). 
 In the latter case, we compare the two procedures mentioned in \S3.2 to describe the thermodynamics
 properties of a mixture of H/He 
 and heavy elements. Comparison between these two limiting cases
should provide the maximum effect on the planet's radius due to the unknown distribution of 
heavy elements within its interior, as long as large-scale adiabatic convection is considered as the heat
transport mechanism.
 All the results portrayed in Fig. \ref{fig2}b  are done with water.
We find the following results:
 \begin{itemize}
\item{} Under assumption (B), models based on a
$\yeff$, for such a high metal fraction, $Z=50\%$, lead to cooling sequences that differ drastically
from the ones based on the AVL, with the SESAME EOS. This shows that the $\yeff$ simplification
can not be used for such values of $Z$, as anticipated from the limitations
of this assumption (see \S3.3).
 \item{} Case (A) and case (B),
 with the AVL and the SESAME EOS, yield similar cooling sequences, with $\sim$ 4\% difference
 at t=1 Gyr (7\% at 5 Gyr). 
 \item{} In case B, we find a 
significant difference when using the AVL
 with ANEOS compared with all other cases ($\sim$ 30\% difference in R for t $\ge$ 1 Gyr compared with case (A),
 see Table 1). As mentioned in \S 3.2 and illustrated in Fig. \ref{figPTS},
 this stems from the different entropy dependence on $(P,T)$ predicted by the different EOS in the relevant pressure regime indicated in the middle panel of Fig. \ref{fig3}.  Although all the evolution sequences are forced to start from the same initial entropy state, the sequence based on the ANEOS EOS shows a much stronger variation of the entropy with time. This stems from the  stronger variation of $S$ with $P$ and $T$ predicted by
 this EOS, for the mass 
 range characteristic of Neptune-like planets (see Fig. \ref{figPTS}). This is highlighted in the upper panel of Fig. \ref{fig3}:  the sequence based on the
 AVL with ANEOS shows $\sim 10$ times larger local gravothermal energy, $-TdS/dt$, at the beginning of the evolution.
 Consequently, this sequence contracts and cools much faster than the other 
ones, loosing its internal entropy at a faster rate (lower panel of Fig. \ref{fig3}) and reaching a significantly smaller radius after 1 Gyr (see Fig. \ref{fig2}b). 

The reason why the difference between the entropy, $S(P,T)$, obtained with the ANEOS and SESAME EOS
(Fig. \ref{figPTS} and Fig. 3) is inconsequential when all the Z-material is located in the core, whereas it yields
significantly different cooling sequences when the Z-component is mixed with the H/He envelope
stems from the fact that the evolution of a planet {\it with a core}
 is dominated by the entropy variation of the H/He envelope, $| dS/dt|_{HHe} \gg   |dS/dt|_{Z}$,
 a consequence of the smaller compressibility, $[\rho (dP/d\rho)]^{-1}$, and smaller specific heat of the Z-material
 compared with the H/He one because of its much larger mean molecular weight (see \S 4.3). 
 \end{itemize}

 Figure \ref{fig2}c also shows the evolution of a model with a more realistic distribution of the heavy material
 in the interior, namely a 8 $\mearth$ core surrounded by a gaseous
 envelope with a $\zenv$=16.6\% heavy element mass fraction, which corresponds to a total
 heavy element enrichment of 50\%. First of all, we note that, for this value of $Z$, the cooling sequence base on the
 $Y_{eff}$ formalism (dash-dot line) is very similar (within $\simle$2\%) to the one based on the more rigorous AVL (long-dash line). This shows that the
$Y_{eff}$ approximation for the treatment of a multispecies H/He/Z EOS can be used relatively
safely up to $Z\approx 20\%$. Second of all, sequences with $M_{core}=8\,\mearth$,
$\zenv$=16.6\% (with the AVL and the SESAME EOS or with a $\yeff$) and the ones with $M_{core}=0$, $\zenv$=50\% (solid line) differ by less than 4\%. 
Therefore,
 for such levels of $\zenv$, the models are less sensitive to the treatment of the Z-element in the envelope. 
 As mentioned previously, the sequence based on
 the ANEOS EOS (short-dash line) yields a significantly faster cooling sequence, with a $\sim 6\%-10\%$ smaller radius at 1 Gyr, compared with all other sequences. This large uncertainty on the EOS will be the major culprit for preventing accurate determinations
 of the exoplanet internal composition from their observed radius. 
 \medskip

We have also explored the effects of the heavy element distribution for smaller total enrichments,
$Z < 15$\%. In this case, we find a $<2\%$ global effect on the radius at a given age, depending whether the heavy elements are all located in the core or are distributed throughout the entire planet, using either an effective He abundance or the AVL with the SESAME EOS for the thermodynamics of the heavy component. 
Models based on the ANEOS EOS (with heavy elements distributed over the entire planet) still
predict the smallest radii at a given time, with a maximum $\sim 10\%$ effect compared with the other sequences.

\begin{figure}
\psfig{file=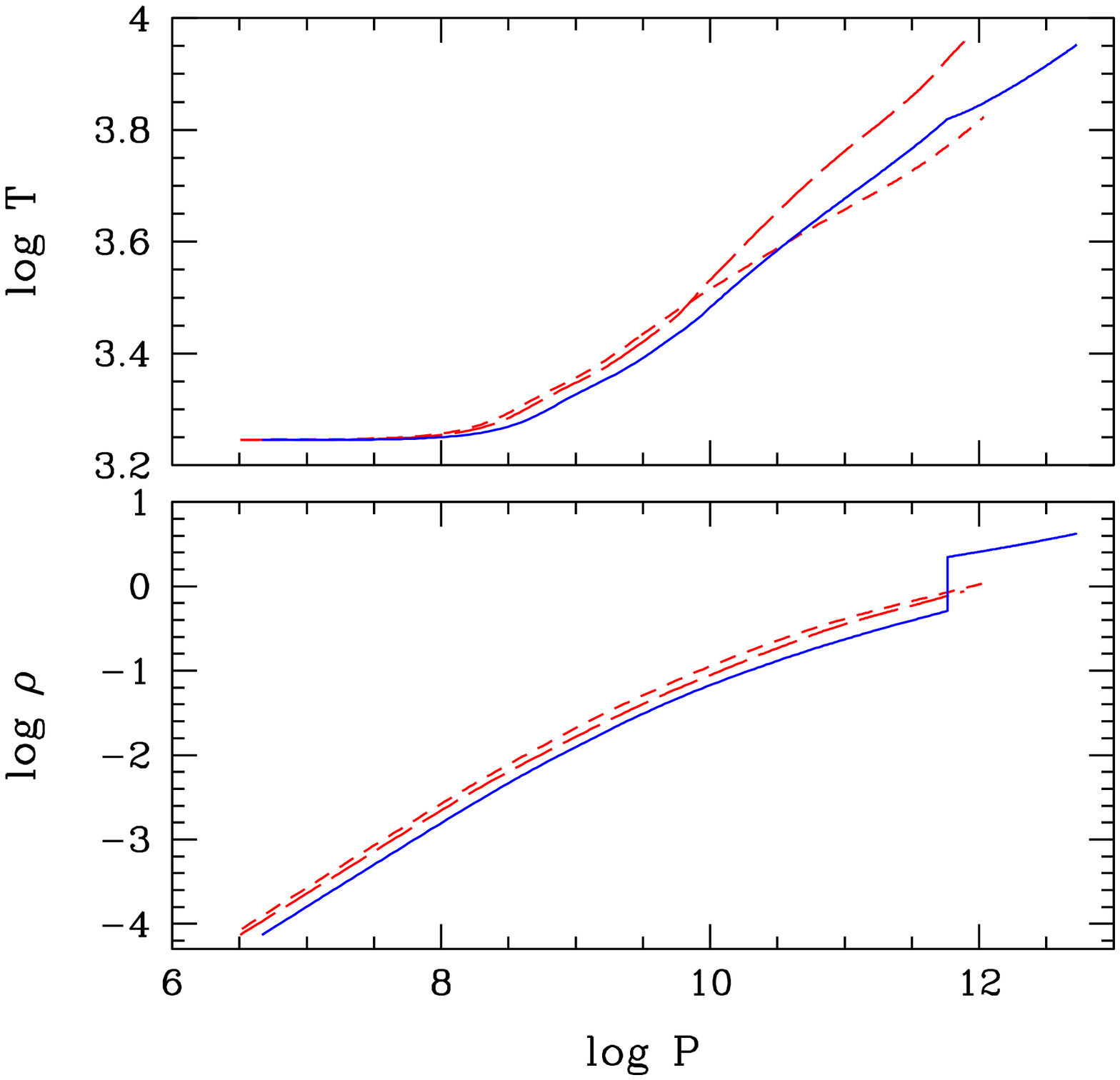,height=110mm,width=80mm} 
\caption{Internal temperature (K) and density (g cm$^{-3}$) profiles for a 20 $\mearth$ planet,
irradiated by a Sun at 0.045 AU, at an age of 4 Gyr.
Solid line: $M_{core}=0.5\, M_{\rm p}$, $\zenv$=0;
long-dash line: $M_{core}=0$, $Z=0.5$, AVL SESAME EOS;
short-dash line: $M_{core}=0$, $Z=0.5$, AVL ANEOS EOS.
}
\label{pro_mj006}
\end{figure}

Figure \ref{pro_mj006} portrays the internal $T(P)$ and $\rho(P)$ profiles for the various models of
our 20 $\mearth$ planet at an age of 4 Gyr.
\medskip
 
The first conclusions to be drawn from these tests for a Neptune-mass planet are the following:

(i) for a core mass less than 50\% of the planet's mass,  a variation of the core composition from pure
water to pure iron, the maximum expected difference in mean densities, yields a
difference on the radius of less than $\sim$ 10\% after 1 Gyr. Yet, such a difference is
 accessible to the observational determination
 of some transit planet radii, {\it if} (when) other sources of uncertainties, in particular on the heavy material
 EOS, were (will be) under control.
  
 (ii) for a metal-fraction in the envelope $\zenv \simle$ 20\%, the EOS of the Z-material
 can be approximated by using a corresponding $\yeff$ effective helium fraction in the SCVH EOS.
 This approximation become more dubious, and even wrong, above this limit.

(iii) if $Z \simle 15$\%, either globally or in the envelope,
the different treatments of heavy elements
yield
a relative variation of the radius of $\simle 2\%$, {\it except when using the ANEOS EOS, which yields
a  $\sim 10\%$ difference}.
Within this limit for Z, and given all the other uncertainties in planet modelling, 
the impact of heavy elements on the evolution of Neptune-mass planets can be mimicked reasonably well
by considering that the heavy elements are all located in the core.
 
(iv) For larger heavy element enrichments ($Z \simgr 15$\%), the distribution (everything in the core versus uniform distribution) and the treatment of 
the heavy element thermodynamic properties ($\yeff$ or AVL) can significantly affect
the cooling and thus radius determination for a given age (more than 10\% difference
after 1 Gyr). These uncertainties, unfortunately, hamper an accurate determination of the detailed composition
 of the heavy element material in the planet's interior.

\subsection{Jovian-mass planets}

\begin{figure}
\psfig{file=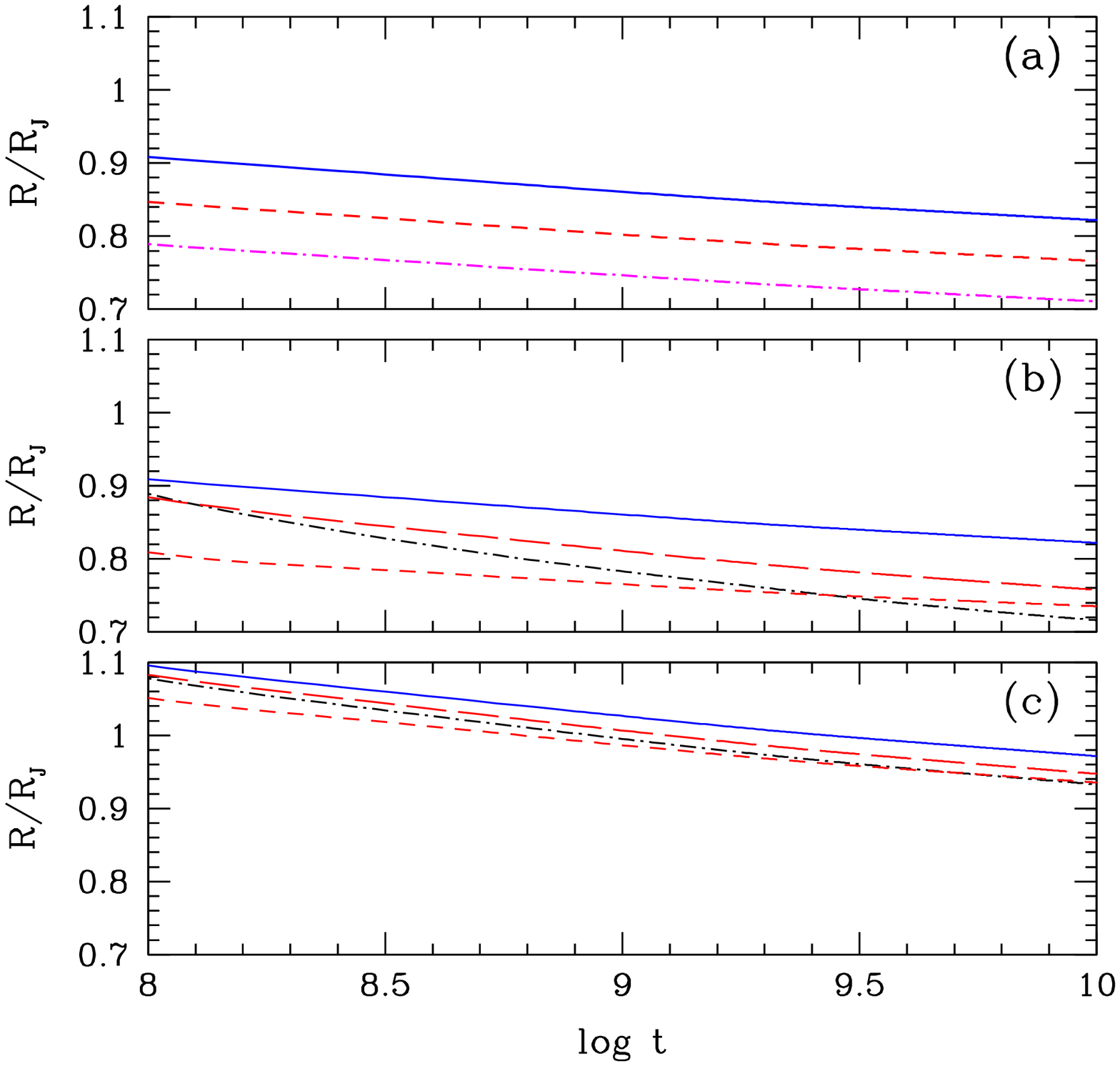,height=110mm,width=80mm} 
\caption{Effect of the composition and the distribution of heavy elements on the radius
 evolution  of a planet of 1 $\mjup$  (318 $\mearth$).
 {\bf (a):} Total heavy element enrichment $Z=50$\%. 
 Models with $\mcore = 159 \mearth$ of water (solid line), rock (dashed-line)
and iron (dotted line), and $\zenv$=0.
{\bf (b):} Total heavy element enrichment $Z=50$\%. The solid line corresponds to a model with $\mcore = 159 \mearth$ of water and $\zenv$=0. The other curves correspond to
models with no core and heavy elements distributed over the entire planet:  mixture of H/He + $Z$ described by  the SCVH EOS with 
$\yeff$=0.275 + 0.5=0.775 (dash-dotted line);  mixture of H/He + water (SESAME EOS)
using the additive volume law (long-dashed line);   
mixture of H/He + water (ANEOS EOS) using the additive volume law (short-dashed line). 
{\bf (c):} Total heavy element enrichment $Z=20$\%. Solid line: model with $\mcore = 63.6 \mearth$ of water and $\zenv$=0; dash-dotted-line: no core and mixture of H/He + $Z$ described by  the SCVH EOS with 
$\yeff$=0.275 + 0.2 = 0.475; long-dashed line: no core and mixture of H/He + water (SESAME EOS) with the AVL and $\zenv$=0.20; short-dashed line: no core and mixture of H/He + water (ANEOS EOS) using the AVL and $\zenv$=0.20.
}
\label{fig4}
\end{figure}

In this section, we extend the analysis done in the previous section
to a template 1 $\mjup$ planet. We first analyse the effect of core composition
for a core mass of 159 $\mearth$, corresponding to a total heavy element enrichment
$Z$=50\% (see Fig. \ref{fig4}a). We find slightly larger effects than for the Neptune-mass case, with a 7\% (15\%) difference in radius between the pure water and the pure rock (iron) core cases,
respectively. Whether such massive cores can indeed form for Jovian-mass planets will be considered in \S7.1. As for the Neptune case, using either the ANEOS or the SESAME EOS for the same material in that case yields similar cooling sequences, with $\simle$3\% differences on $R$ for water and
$\simle$2\% for rock. For these planets, the iron core does not undergo a phase transition
and iron always remains liquid, according to the ANEOS EOS. Indeed, central temperatures
and densities are significantly larger than for Neptune-mass planets, with $T_{\rm c} \sim 10^5$ K and $\rho_{\rm c} \sim 157$ g.cm$^{-3}$ at 1 Gyr for a 1 $\mjup$ planet with an iron core (see Table 2).

\begin{table*}
\caption{
Radius and central thermodynamic properties of a planet of mass $M_P=1 \mjup$ at 1 Gyr for two 
 heavy element mass fractions $Z=(M_Z/M_P)$=50\% and $Z=20$\%. The labels are the same as in Table \ref{tab1}. As mentioned at the end of \S 4.2, for $Z$=50\%, the solution with $\mcore$=0 and $Z=\zenv$
 is similar to a solution with $\mcore \sim 10 \mearth$ and the rest of the heavy material distributed in the
 envelope.}
\label{tab2}
\begin{tabular}{cllllllll}
\hline
\hline 
 $Z$ &$\mcore$ & EOS (core) & $\zenv$ & EOS(env) & R$_{\rm p}$ & $T_{\rm c}$  & $\rho_{\rm c}$  \\
  &  ($\mearth$) &   & & env. & ($\rjup$) & (K) & (g cm$^{-3}$) \\
 \hline
 \hline
0.5 &159 & W-a & 0 & SC &  0.861 &  5.3 10$^4$ &   20.89\\
&159 & W-s & 0 & SC &  0.841 &  6.8 10$^4$ &   28.31\\
&159 & R-a & 0 & SC & 0.802 &  7.2 10$^4$ &   47.11\\
&159 & R-s & 0 & SC & 0.789 &  8.6 10$^4$ &   59.87\\
&159 & I-a & 0 & SC & 0.746 &  1.1 10$^5$ &  156.93\\
&0 &  & 0.5 & SC (0.775$^a$) & 0.782 &  7.7 10$^4$ &    8.88\\
&0 &    & 0.5 & SC+W-a & 0.765 &  3.4 10$^4$ &    8.13\\
&0 &      & 0.5 & SC+W-s & 0.811 &  5.8 10$^4$ &    7.36\\
&&&&&& \\
0.20 &63.6 & W-a & 0 & SC & 1.026 &  4.3  10$^4$ &   13.98\\
&0 &  & 0.2 & SC (0.475$^a$) & 0.994 &  3.9 10$^4$ &    4.38\\
&0 &    & 0.2 & SC+W-a & 0.986 &  3.2 10$^4$ &    4.18\\
&0 &      & 0.2 & SC+W-s & 1.006 &  3.7 10$^4$ &    4.06\\
 \hline
 $^a$ Value of $\yeff$.
 \end{tabular}
\end{table*}

\begin{figure}
\psfig{file=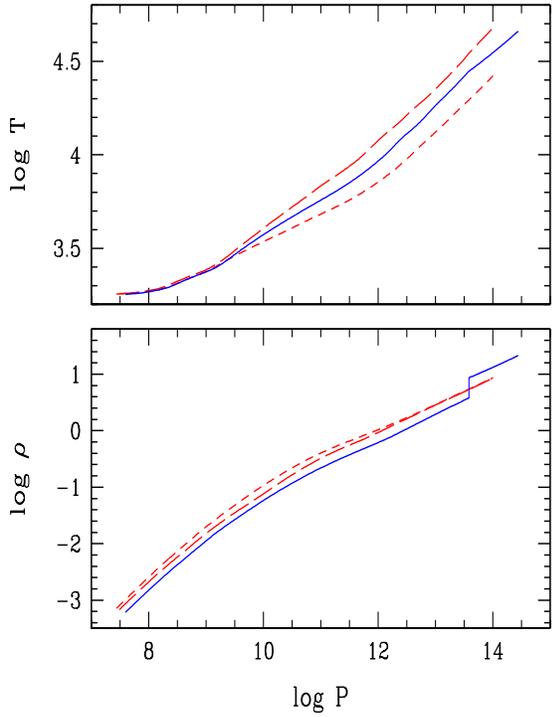,height=110mm,width=80mm} 
\caption{Internal temperature and density profiles for a 1 $M_J$ planet at an age of 4 Gyr.
Solid line: $M_{core}=0.5\, M_{\rm p}$;
long-dash line: $M_{core}=0$, $Z=0.5$, AVL SESAME EOS;
short-dash line: $M_{core}=0$, $Z=0.5$, AVL ANEOS EOS.
}
\label{pro_mj1}
\end{figure}

We explore the effects of the heavy element distribution with $Z=50$\% (Fig. \ref{fig4}b) and $Z=20$\% (Fig. \ref{fig4}c). The latter case is comparable to the expected level of enrichment in Jupiter and Saturn
(Saumon \& Guillot 2004). In the $Z=50$\% case, both the various thermodynamic treatments and
localizations of the heavy elements 
yield significantly different
cooling behaviors (see Fig. \ref{fig4}b). The sequence based on the $\yeff$ approximation, notably,
differs from the other ones, as expected from our previous study for Neptune-mass planets.
At 1 Gyr, the radii obtained with models based on the AVL with ANEOS and SESAME, respectively, differ
by $\sim 6\%$.
The different distributions of heavy elements within the planet have an even larger impact, with up to  12\% difference in $R$ between the case with 
$\mcore$=159 $\mearth$, $\zenv$=0 and the case with $Z$ distributed throughout the entire planet
with ANEOS. Interestingly enough, for the present Jovian conditions,
models based on the ANEOS or SESAME EOS, when metals are mixed throughout the entire planet, yield
smaller differences than for the Neptune-mass planet case, after about 1 Gyr, i.e. at high
pressure and moderately high temperature. Indeed, the inner pressure and temperature 
conditions are very different between Neptune-mass and Jupiter-mass planets (see Figures \ref{pro_mj006} and \ref{pro_mj1}). For the former ones, $P$ ranges between 10$^{10}$ and 10$^{12}$ dyne.cm$^{-2}$ over more than 99\% of the total mass (see Fig. \ref{fig3}) while $T$ ranges from $\sim$ 5000 to 10$^4$ K. For the latter ones, the typical domains are $P=10^{12}-10^{14}$ dyne.cm$^{-2}$ and $T=10^4-10^5$ K. As mentioned in \S 3.2, differences in the $P$ and $T$ dependence of the entropy between the different EOS are more pronounced under the interior conditions of Neptune-mass planets than for Jupiter-mass planets (see Fig. \ref{figPTS}).

If, instead of the extreme $M_{core}=0$ or $Z_{env}=0$ cases, we take a more realistic model
with a 10 $\mearth$ core, comparable to what is expected in Jupiter or Saturn, and we distribute
the rest of heavy elements homogeneously in the H/He rich envelope, we find essentially the same evolution as when the heavy elements are distributed throughout the whole planet, with no core. This suggests that for {\it massive, metal-rich planets},
the evolution is better described by models which assume that all metals are distributed over the {\it entire planet}, since this yields results similar to the ones obtained with a more realistic distribution, than
by models which assume that all heavy elements are in the core, with a metal-free, $Z=0$ envelope.
 
For a more moderate heavy element enrichment, $Z=20$\% (see Fig. \ref{fig4}c), the treatment of heavy elements in the entire planet, based on an $\yeff$ or  on the AVL with ANEOS or SESAME,
is found to be less consequential, for the
present Jupiter-mass planet case. The different methods to describe the EOS yield
 less than 2\% differences on $R$ at a given age. The effect of the distribution of heavy elements
(all in the core versus all distributed over the entire planet) is slightly more consequential,
with up to 4\% difference on $R$. 

Figure \ref{pro_mj1} portrays the internal $T(P)$ and $\rho(P)$ profiles for the various models of
our 1 $M_J$ planet at an age of 4 Gyr.
\medskip
 
The conclusions derived from these tests for a Jupiter-mass planet can be summarized as follows:

(i) for a core mass less than 50\% of the planet's mass,  a variation of the core composition between pure water and pure rock (iron) yields an effect on the radius of $\simle$ 7\%  (15\%).

(ii) for a global metal-enrichment $Z \simle 20$\%, the treatment and distribution of heavy elements
affect only modestly the radius predictions (less than 4\% effect on $R$ at a given age). 
 
(iii) For significant heavy element enrichment ($Z \sim 50$\%), the distribution of heavy elements (everything in the core versus uniform distribution) has a non negligible effect on the radius
predictions, up to 12\% difference at a given age. For such high metal-enrichments, distributing all the
heavy material throughout the entire planet, with no core, is found to yield similar results as a more
realistic model with about a 10 $\mearth$ core and the rest of the material in the envelope, while a
model with a metal-free envelope and all heavy elements in the core differs substantially from these sequences.


\subsection{Energy release from the core}

As mentioned in \S 2.2, the thermodynamics of the heavy material component in planet cores is often described either by a
zero-temperature EOS or
under the assumption of a uniform, low temperature (Fortney et al. 2007; Seager et al. 2007),
or a uniform density (Bodenheimer et al. 2003). These models ignore the gravothermal energy
contribution of the core, $(-T {dS \over dt})_{\rm core}$ to the cooling of the planet. As an other extreme,
Burrows et al. (2007)
arbitrarily assume that the specific entropy of the core material is the same as that of the H/He envelope, therefore
overestimating the heat release of the dense core.


According to the ANEOS or SESAME EOS, the typical heat capacity of water in the cores of
the presently studied Neptune-mass and Jovian-mass planets is  
$C_{\rm v}\sim 3\times 10^7$-5$\times 10^7$ erg g$^{-1}$ $K^{-1}$, for typical temperatures $ T \sim 5000$-5$\times 10^4$ K and pressures $P \sim 10^{12} - 10^{14}$ dyne cm$^{-2}$ (see Figs. \ref{pro_mj006} and
\ref{pro_mj1}). This is about 
1/3 the specific heat  of H/He in the envelope. This corresponds to variations of the specific heat
from the high-temperature nearly ideal gas limit,
$\sim 6{\mathcal R}/A $ (where
$A$ is the species mean atomic weight and ${\mathcal R}$ is the perfect gas constant), to regimes which include
 potential-energy contributions associated with translational and librational modes ($3{\mathcal R} /A $ maximum for three translational and three librational modes per molecule).
If the species enters the solid phase, the specific heat decreases rapidly (Debye regime) and eventually vanishes.
The specific heat of rock or iron is even smaller (larger atomic mass) and the contribution of cores made of such materials to the cooling history of planets is negligible. 
For planets in the mass range 20 $\mearth$ - 1 $\mjup$, as explored in the previous section,
with core masses less than 50\% of the planet's mass, and with cores
made up of water, the release of gravitational energy 
($P {dV \over dt}$) and of internal energy (${dU \over dt}$) never exceeds  40 \% of the
total released energy, $T ({dS \over dt})$. 
For Neptune-mass planets, the dominant contribution from the core is due to its contraction, i.e the $P {dV \over dt}$ work, during
the first 0.5 - 1 Gyr of evolution, and by the release of its internal energy at older ages. For jovian-mass
planets, the contraction of the core provides the dominant contribution to its gravothermal energy during
the entire planet evolution. 

The assumption of zero-T or uniform low temperature affects  
the structure of the cores, because of the temperature dependence of the $P(\rho$)
relations predicted
by current EOS (see \S 3.2 and Fig. \ref{fig1}).
But, more importantly, it affects the evolution of the planet, because it 
implies a negligible contribution of the core to the total gravothermal energy of the planet. Indeed,
the release of thermal energy from the core, $({dU \over dt})_{\rm core}$, is forced to be zero, and the small compressibility of water at low T
drastically underestimates the true contraction of the core during the planet evolution, and thus the release of
gravitational energy.
We have analysed this assumption by  imposing a constant and uniform temperature of 300 K in 
the water core of the planet models analysed in the previous section, using ANEOS and SESAME.  
For the jovian-mass planet (1 $\mjup$) with $\mcore$=159 $\mearth$, the effect on the radius is negligible (less than 2\%) after 1 Gyr. On the Neptune-mass planet (20 $\mearth$),
the effect is larger, the largest effect being found with the SESAME EOS for the model with a
10 $\mearth$ core, that yields
a 6\% effect on $R$ after 1 Gyr. The largest temperature variations of the EOS are indeed found at low pressure ($P < $ 1 Mbar), as illustrated in Fig. \ref{fig1}, affecting more importantly light planets.

The heat transport in the core is due to convection or conduction
(by electrons or phonons), depending on the core material and the age of the planet. Conduction may dominate after a few Gyr of evolution, as the core becomes cool and dense enough for the thermal conductivity due to degenerate electrons to become large enough, $K_c\propto \rho^{4/3}$,
and the conductive flux to dominate the convective one, ${\mathcal F}_{cond}=-K_c\nabla T>{\mathcal F}_{conv}$. It starts to dominate earlier for rock and iron cores, compared to water cases. These results,
however, are hampered by the uncertainties in the conductive opacities calculated with the
ANEOS EOS or with Potekhin (1999) for such materials. We have checked the effect of such an uncertainty
on the cooling history 
in the case of the 20$\mearth$ planet with  $\mcore$=10 $\mearth$ of water with the SESAME EOS, as  this sequence provides a case with the largest energetic contribution from the core, for enrichments $Z \le $ 50 \%. Over the entire evolution, heat is predicted to be transported by convection within the core and the temperature gradient is given by the adiabatic gradient. We have arbitrarily decreased the conductive opacities so that conduction now dominates over convection. The core thus becomes isothermal. The effect remains small on the radius evolution (maximum 3\% on $R$ at a given age compared to the  convective case). Therefore, the uncertainty in the heat transport
efficiency of the core has a smaller impact than neglecting the temperature dependence of the core material and its global energetic contribution.

To conclude this section, the present calculations show that neglecting the thermal and gravitational contributions
from the core, by assuming zero-T or low uniform T,  in current planet modelling,
 leads to a maximum effect of $\sim$ 6\%  on the radius, after 1 Gyr of evolution,
 for cores less massive than 50\% of the planet's mass. Whether such a variation can
 be considered negligible or not depends on the accuracy of the data the models need to be
 confronted to in order to infer the planet's internal composition, once the EOS of heavy elements
 will be determine with enough accuracy.
 As will be shown in the next section, the
conclusion is drastically different for larger metal contents, for Neptune-size planets.

\section{Extreme metal-enrichment: the two specific cases of  HD149026b and GJ436b}

In this section, we analyse extreme cases of heavy element enrichments ($Z > 50$\%), focusing on two recently discovered transiting planets, the Saturn-mass planet HD149026b and the Neptune-mass planet GJ436b.
Our goal is to examine the properties of these two transit planets, in the light of the analysis
conducted in the previous section concerning the main assumptions and uncertainties of current planetary models.

\subsection{HD149026b}

The discovery of the Saturn-mass planet HD149026b (Sato et al. 2005)
revealed an unexpectedly dense planet, with a mass $M_{\rm P}=0.36 \pm 0.04 \,\mjup$ and
a radius $R_{\rm P}=0.725 \pm 0.05\, \rjup$, i.e. a mean density $\bar \rho =1.17\pm 0.35$ g.cm$^{-3}$. For comparison, Saturn has a mass
of 0.3 $\mjup$ but a radius of $\sim$ 0.8 $\rjup$, i.e. $\bar \rho =0.66$ g.cm$^{-3}$.  The transit planet is orbiting
a G-type star at an orbital distance $a=0.042$ AU, about 230 times closer to its Sun than Saturn. The age of the system, $\sim$  ($2 \pm 0.8$) Gyr, is also younger than our Solar System. 
It is important to understand the nature and the origin of this puzzling planet, and in particular
to know whether  current planet formation scenarios, in particular the core accretion model, can explain it. This requires a knowledge of its structure and composition, which can only be inferred from theoretical models. Several authors have tried to infer the inner structure of this planet under the usual assumptions described in \S 2. The models  assume that almost 
all heavy elements are in the core,
and the H/He envelope is either free of metals (Burrows et al. 2007), 
or is moderately enriched, with $Z=Z_\odot$ (Ikoma et al. 2006) or $Z=0.045$ (Fortney et al. 2007).
Ikoma et al. (2006) have also investigated a case with $\zenv=0.37$.
  Fortney et al. (2007) use a zero-temperature EOS for the core and thus ignore its 
heat content contribution. They find negligible effect when using a nonzero temperature EOS for
the ice mixture given by Hubbard \& Marley (1989). This latter EOS provides pressure-density relations  appropriate for the description of warm adiabatic mixtures ($T \sim 10^4$) K, but do not explicitly account for the temperature dependence. Ikoma et al. (2006) also use this EOS and assume
 a uniform temperature in the core and a specific heat $C_{\rm v} \sim 10^7$ erg g$^{-1}$ $K^{-1}$ to account for the core heat release.
According to all these models, the total mass of heavy elements in the planet lies in the range $\sim$ 40-90$\mearth$, {\it i.e} $Z \sim$ 35\% - 80\%. 

\begin{figure}
\psfig{file=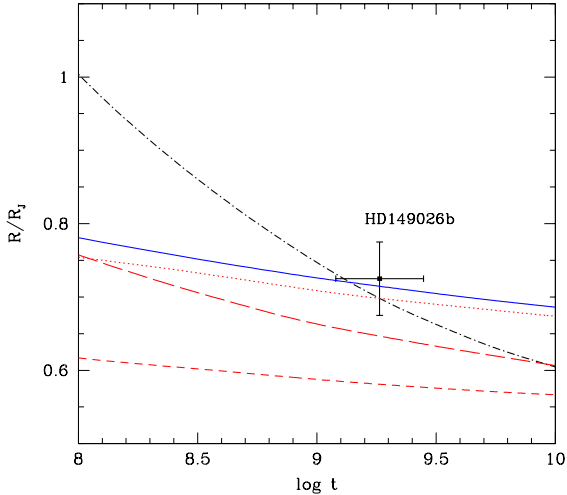,height=80mm,width=80mm} 
\caption{Evolution of a planet with the characteristics of HD149206b ($m_{\rm P}$=116 $\mearth \sim 0.36 \mjup$)
with different heavy element distributions. 
Solid line: model with $\mcore = 80 \mearth$ of water (ANEOS EOS) and $\zenv$=0. 
The other curves correspond to
models with  heavy elements distributed over the entire planet.
Dash-dot line: mixture of H/He + $Z$ described by  the SCVH EOS with an effective helium
abundance, $\yeff$=0.275+0.69=0.965, which corresponds to $Z=0.69$,
{\it i.e} $80 \mearth$ of heavy elements. Long-dash line: mixture of H/He + water (SESAME EOS) using the AVL with $Z$=69\%.  Short-dash line: 
mixture of H/He + water (ANEOS EOS) using the AVL with $Z$=69\%. 
Dotted line: Mixture of H/He + water (ANEOS EOS) using the AVL with $Z$=51\%, {\it i.e} 60 $\mearth$ of heavy elements.}
\label{fig5}
\end{figure}

We will test the impact of these main assumptions done in current structure models of HD149026b,
namely the distribution of all heavy elements in a core and  the use of zero-T EOS.
Since the present study focusses on the sensitivity of the
structure and evolution for a given setup of atmosphere models, we use, as outer boundary condition to the inner structure, atmosphere models with solar composition, $Z_\odot$, irradiated by a G-star at 0.04 AU, even though the results are expected to change to some extent with the composition of the atmosphere (see e.g Burrows et al. 2007). This issue will be explored in a forthcoming paper.
 We obtain a good match of the planet's radius with a model with a water core of 80 $\mearth$ and
a metal-free envelope, $\zenv=0$. Figure \ref{fig5} displays this model (solid line), calculated with the ANEOS EOS in the core. The use of the SESAME EOS for water  in the core yields a 3\% smaller radius,
still providing a good match to the observed value. Figure \ref{fig5} also shows the impact
 of the heavy element distribution, with models where the 80 $\mearth$ of heavy elements are
 distributed 
over the entire planet.
As anticipated from the studies conducted in \S 3.2, the use of an $\yeff$ in the SCVH EOS to handle the heavy element contribution yields, for such a high metal fraction, an evolution that differs drastically from the other ones, even though,
coincidentally, it gives a good solution {\it at the age of the system}. As expected from the tests performed in \S 4, models with 80 $\mearth$ of heavy elements
mixed with H/He throughout the entire planet, using the AVL with the SESAME
 (long-dash line) or the ANEOS (short-dash line) EOS
for water, yield denser structures. The latter one yields a radius $\sim$ 25\% smaller than the observed value, an effect larger than
changing the core composition from water to rock ($\sim$10\% effect on $R$),
for the same core mass of 80 $\mearth$, and similar to the one obtained when changing from a pure water to a pure {\it iron} core, a rather unlikely solution.
 
 Instead of the two aforedescribed extreme heavy element distributions,
we have also examined a more realistic model with a 20 $\mearth$ core, similar to what is predicted for Saturn,
and the remaining 60 $\mearth$ mixed within the H/He envelope, using the AVL both with SESAME
and with ANEOS.
 As anticipated from the tests performed in \S 4.2, this yields similar results than assuming that all heavy elements are distributed throughout the entire planet. Models with such metal enrichment
 and distributions thus seem to be excluded by the observations.
An alternative solution for HD149026b, however, is  obtained with a
 60 $\mearth$ of heavy elements, instead of 80, mixed with H/He throughout the entire planet,
 described with the ANEOS EOS for water and the AVL formalism (dotted line in Fig. \ref{fig5}),
 whereas when using the SESAME EOS for water, 70 $\mearth$ are required.
 As mentioned above and as shown in our previous studies, such models are equivalent to models
 with a small core mass and a significantly enriched envelope. 
This shows that for a given heavy element material, water in the present case, the effect of modifying its internal distribution (everything in the core or a fraction redistributed in the H/He envelope) has
by itself a large impact, yielding an uncertainty on the amount of heavy material required to reproduce
the observed radius of 80-60=20 $\mearth$. As shown in the previous sections,
assuming all heavy elements to be in the core yields structures less dense than when these elements are mixed
throughout the envelope, so that these models require a larger amount of heavy material to match
a given radius.
 
Finally, as done in \S 4.3, we have tested the effect of the temperature dependence of the EOS in the core by assuming a uniform temperature of 300 K in a model with a 80 $\mearth$ water core, using ANEOS and SESAME EOS. 
The largest effect is found with this latter EOS, the model with a uniform low temperature yielding
a radius $\sim$ 3 \% smaller than the "hot" core case at 2 Gyr.


\subsection{GJ436b}

The first Neptune-mass transiting planet has been discovered recently (Gillon et al. 2007a),
with a mass $M_{\rm P}=22.6 \pm 1.9 \mearth$ and a radius
$R_{\rm P} = 25200 \pm 2200$ km = 0.352 $\pm 0.03\, R_{\rm J}$. 
The planet is orbiting an M-type star of $\sim$ 0.44 $\msol$ at an orbital distance $a \sim 0.028$ AU. According to
the models of Fortney et al. (2007), Gillon et al. (2007a) suggest that the planet is
composed predominantly of ice with a thin H/He envelope of less than 10\% in mass.
Based on Spitzer observations, Gillon et al. (2007b) and Deming et al. (2007) determine a slightly larger radius, with $R_{\rm P}$=0.386 $\pm 0.016 \rjup$. 
We adopt this value in the following and we assume an age for the system of 1-5 Gyr, as it is essentially unconstrained by the observations.
Models available in the literature to determine the inner structure
of this planet use temperature-independent EOS to describe the core and thus ignore its thermal contribution (Fortney et al. 2007; Seager et al. 2007; Adams et al. 2007). We will test
this assumption.

\begin{figure}
\psfig{file=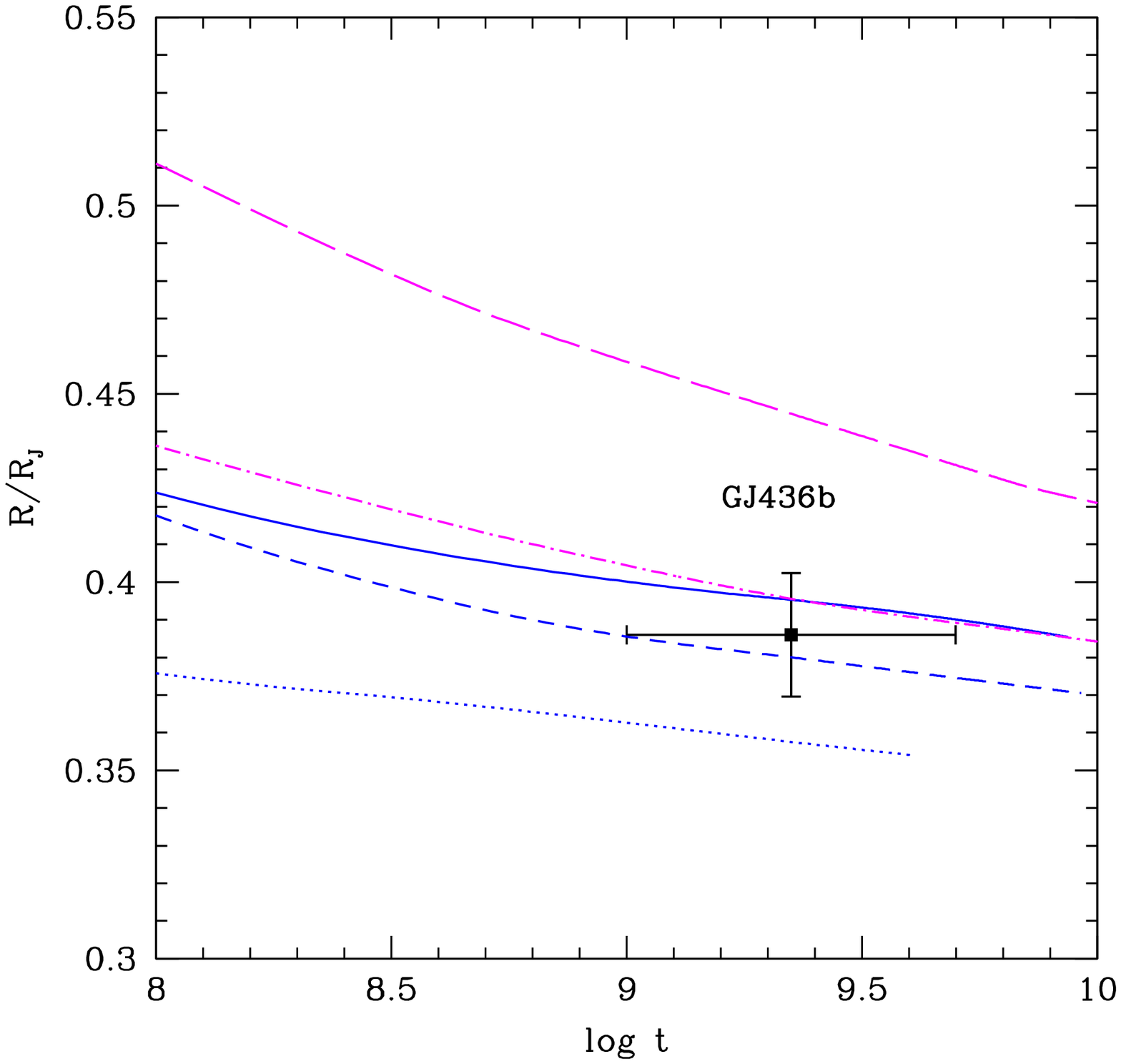,height=80mm,width=80mm} 
\caption{Evolution of a planet characteristic of GJ436b ($M_{\rm P}$=22.6 $\mearth$,)
with different heavy element distributions. Models with no irradiation:
Solid line: model with $\mcore = 21 \mearth$ of water (ANEOS EOS) and $\zenv$=0;
short-dash line: model with $\mcore = 21 \mearth$ of water (SESAME EOS) and $\zenv$=0;
dotted line: test model with $\mcore = 21 \mearth$ of water (SESAME EOS) and $\zenv$=0,
assuming an uniform temperature of 300 K in the core. 
Effect of irradiation, with $F_{\rm inc}=6\times F_{\rm inc}$ for GJ436b (see text):
long-dash line: same as short-dash line with irradiated atmosphere models;
dash-dot line: same as dotted line with irradiated atmosphere models
}
\label{figGj436}
\end{figure}

We have calculated models characteristic of GJ436b.  We use solar metallicity atmosphere models and, given the low luminosity of the parent star, we neglect presently the irradiation effects. Recent determinations of
GJ436b irradiation induced temperature (i.e. brightness temperature) indeed suggest that the evolution is not
likely to be significantly altered by irradiation (Demory et al. 2007).  
A model with a core of 21 $\mearth$ made of water
and a metal-free envelope, $\zenv$=0,
provides a good match to the observed radius, as illustrated in  Fig. \ref{figGj436} (solid and dashed lines).
The ANEOS EOS for water (solid line) yields
a slightly larger ($\sim$ 4\%) radius than the SESAME EOS (dashed line).
A good match is also obtained with a rocky (dunite) core of mass 19.5 $\mearth$.
Given the high mass fraction of heavy material  of this planet, more than 80\%, the freedom to vary
its distribution is limited.
Assuming a core of 21 $\mearth$ with $\zenv$=0 or a slightly smaller core with some heavy element enrichment in the envelope, a more realistic solution, e.g. a model with $\mcore$ = 20 $\mearth$ and $\zenv$=0.38, yield less than 8\% variations on the radius, still within the observed error bars.
According to the tests performed in \S 4.1 on a 20 $\mearth$ planet,
larger values of $\zenv$ ($\zenv \, \simgr$ 50\%), and thus smaller cores, yield larger effects on the radius. Note that, according to present planet formation models (Alibert et al. 2005a; Mordasini et al. 2008), such high fractions of Z-material enrichment in envelopes of Neptune-mass planets are
not excluded (Baraffe et al. 2006, figure 1) since, as mentioned in \S 2.1, the fate of accreted planetesimals during the gas accretion phase depends on the envelope mass.

\begin{figure}
\psfig{file=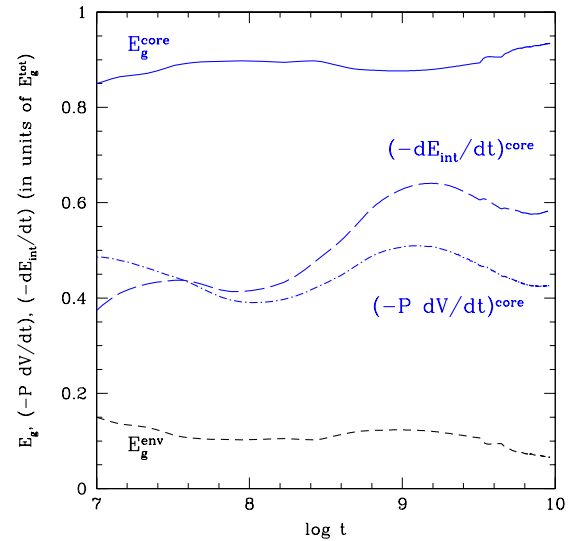,height=80mm,width=80mm} 
\caption{Various contributions to the gravothermal energy, $E_g=-TdS/dt$, normalized to the total value, for a planet characteristic of GJ436b ($M_{\rm P}$=22.6 $\mearth$), with a 21 $\mearth$ core of water with the SESAME EOS.  Solid line: global contribution from the core; dash-dot line: contraction
work contribution from the core; long-dash line: thermal contribution from the core; short-dash line:
global contribution from the H/He 1.6 $\mearth$ envelope.
}
\label{figteg}
\end{figure}

Figure \ref{figGj436} also displays the evolution of the aforementioned model with the assumption
of a uniform cold temperature (300K) in the core. This yields a moderate effect on the evolution,
about 4\% on the radius  at 1-5 Gyr with the ANEOS EOS and $\sim$ 6\% effect with SESAME.
In the latter case, however, the model lies outside the observational error bars (see dotted line in Fig. \ref{figGj436}).
As seen in Fig. \ref{figteg}, for planets with such a large fraction of heavy material, and conversely
with such a modest gaseous H/He fraction ($\simle 10\%$), the evolution of the planet
is largely dominated by the core contribution. Neglecting the temperature-dependence of the core, and thus
its contribution to the gravothermal energy  of the planet,
yields an incorrect cooling sequence.
Even though, for such low-entropy material, the global gravothermal energy, $TdS/dt$, remains small, a proper calculation should take into consideration the core contribution. Note also that even a modest H/He fraction affects the radius determination, as illustrated in Table \ref{tab3} by comparing the present results with
the value of the radius corresponding to a pure 22.6 $\mearth$
icy planet with no gas envelope, as derived from the fitting formulae of Seager et al. (2007).

\begin{table}
 \caption{Radius of a 22.6 $\mearth$ planet at 2 Gyr with different water core masses, EOS and levels of irradiation, characterised by a stellar incident flux, F$_{\rm inc}$. Two cases are considered: no irradiation (F$_{\rm inc}$=0)
and F$_{\rm inc}$=6 x F$_{\rm inc}$ of GJ436b. Comparison is done with the pure water (ice) case of Seager et al. (2007).
Results based on the assumption of a uniform core temperature of T=300K are also given.}
\begin{tabular}{llll}
\hline
\hline 
 $\mcore$ & F$_{\rm inc}$ & EOS (core) & R$_{\rm p}$   \\
    ($\mearth$) &   & &  ($\rjup$) \\
 \hline
 \hline
21 & 0 & aneos & 0.396 \\
21 & 0 &  aneos T=300K & 0.379 \\
21 & 0 & sesame & 0.381\\
21 & 0 & sesame T=300K & 0.358 \\
22.6 & 0 & Seager & 0.285 \\
21 & 6 x F$_{\rm inc}^{\rm GJ436}$ & aneos & 0.452 \\
21 & 6 x F$_{\rm inc}^{\rm GJ436}$ & aneos T=300K &0.421  \\
21 & 6 x F$_{\rm inc}^{\rm GJ436}$ & sesame & 0.446 \\
21 & 6 x F$_{\rm inc}^{\rm GJ436}$ & sesame T=300K & 0.397 \\
 \hline
  \end{tabular}
  \label{tab3}
\end{table}

Importantly enough, taking into account the thermal and gravitational energetic contributions from the core becomes even more
crucial if the irradiation effects from the parent star are important. Fig. \ref{figGj436} shows the evolution of  the same planet model, with $\mcore = 21 \mearth$ of water (SESAME EOS), but with
an incident stellar flux $F_{\rm inc}= {1 \over 2} ({R_\star \over a})^2 F_\star$,  where $F_\star$ is the flux from the parent star, six times larger than for GJ436a. This corresponds to a parent star about 50\% hotter.
In that case, neglecting the temperature dependence of the EOS in the core and its
contribution  to the planet's cooling yields
a $\sim$ 11\% (7\%) smaller radius with SESAME (ANEOS) EOS (dash-dot vs long-dash curves). This stems from the larger planet
interior temperature and entropy in the irradiated sequence compared with the non-irradiated one.
In the irradiated case, the core temperature ranges from 5000 to $\sim 2\times10^4$ K (from the bottom of the H/He envelope to the center),
to be compared with 3000 to $\sim10^4$ K in the non-irradiated case,
characteristic of the temperatures
expected in the rocky/icy part of Neptune and Uranus (Guillot 2005).


\section{Evolution of super Jupiter planets: Hat-P-2b and deuterium burning planets}

\begin{figure}
\psfig{file=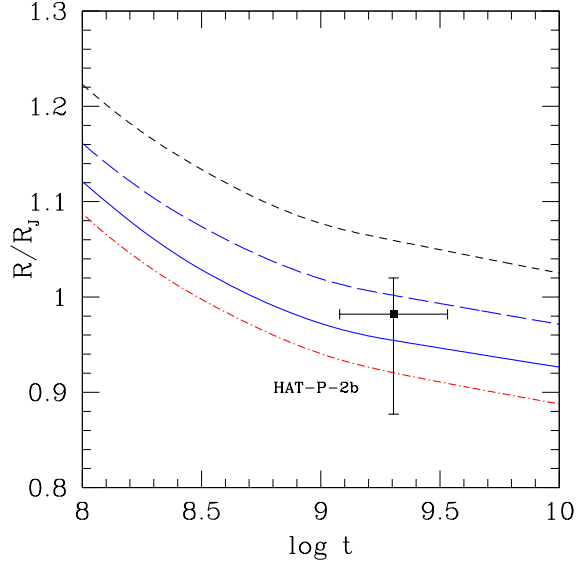,height=80mm,width=80mm} 
\caption{Evolution of a super-Jupiter planet with the characteristics of HAT-P-2b 
($M_{\rm p}$= 9 $\mjup$)
for different distributions of heavy elements. Irradiation effects from the parent star are taken into
account, adopting a time-average orbital distance $a=0.077$ AU, for the proper eccentricity $e$=0.52
and semimajor axis of the relative orbit $a_{\rm rel}$=0.0677 AU.
Solid line: $\mcore = 600 \mearth$ of water (ANEOS EOS) and $\zenv$=0.
Long-dashed line: $\mcore = 350 \mearth$ of water (ANEOS EOS) and $\zenv$=0.
Dash-dotted line: 600 $\mearth$ of heavy elements with 100 $\mearth$ in the core (water,  with ANEOS EOS) and 500 $\mearth$ in the envelope, i.e. $Z_{env}=0.18$, mimicked by an $\yeff$=0.275 + 0.18.
Short-dashed line: H/He brown dwarf with $Z=Z_\odot$.}
\label{fig7}
\end{figure}

The final part of our study is devoted to "super-Jupiter" extra-solar planets, 
with masses $M_{\rm P}\gg$ 1 $\mjup$.  It is motivated by the growing number of discoveries
of massive extra-solar planets, in the mass 
regime overlapping the one of low-mass brown dwarfs, issued from a different formation mechanism.
These discoveries  feed the heated debate concerning the definition
of a planet and the possibility to distinguish planets from brown dwarfs of similar mass. 
In this context, one of the most remarkable discoveries is the transiting super-Jupiter object HAT-P-2b
(also named HD147506b), with a mass  $M_{\rm p}$ = 9.04 $\mjup$ and 
a radius $R_{\rm p}$ = 0.982 $\rjup$ (Bakos et al. 2007).
Loeillet et al. (2007) reanalysed the orbital parameters of the system and 
find 
similar values,  $M_{\rm p}$ = 8.64 $\mjup$, $R_{\rm p}$ = 0.952 $\rjup$. In that case, the mass-radius relationship offers a unique information to infer the gross composition of the object and to determine
its real nature, low-mass gaseous brown dwarf or very metal-rich massive planet.
Bakos et al. (2007) suggest that the mean density of this planet is only marginally consistent
with model predictions for an object  composed predominantly of H and He, and
requires the presence of a large core, with $\mcore \simgr 100 \mearth$. Here, we calculate more thoroughly the internal structures consistent with the radius determination, along the lines described in the previous sections.
We find that a total amount of $\sim$ 300-600\,\,(200-500)$\mearth$ of a water (rock) component is required
to explain the radius at the present age, as illustrated in Fig. \ref{fig7}. 
The rather large predicted range of heavy material enrichment stems from the large observational error bars.
As seen in the figure, a brown dwarf (Baraffe et al. 2003), i.e. 
a gaseous H/He object with
solar metallicity, is predicted to have at this age a radius marginally consistent at the 2$\sigma$ limit with the observations (short-dashed line) and thus can be excluded at the $\ga 95\%$ confidence level, if
the present radius observational determination is confirmed. This shows that
planets
can exist with masses up to $\sim$ 10 $\mjup$, well above the opacity-limit for fragmentation,
$m\sim 3$-5 $\mjup$ (\cite{Whitworth06}), the expected minimum mass for brown dwarf formation.
These two distinct
astrophysical populations should then overlap over a substantial mass domain.

In this context, it is interesting to
explore the fate of
{\it planets} massive enough to ignite deuterium-fusion in their central parts, i.e. with
$M_{\rm p} \simgr 12\, \mjup$ (Saumon et al. 1996, Chabrier et al. 2000). 
 Indeed, recent calculations of planetary population synthesis based on improved core-accretion models of planet formation
(Alibert et al. 2005a) and a wide variety of initial conditions, predict the formation of super-massive
planets, up to $\sim$ 20-25 $\mjup$, with rocky/icy core masses up to several $100$ $\mearth$ (Ida \& Lin 2004; Mordasini et al. 2007, 2008). If planets with a massive core can form above the
aforementioned deuterium burning minimum mass, a key question is to determine whether or not the presence of the core can prevent
deuterium burning to occur in the deepest layers  of the H/He envelope. Guided by the results of Mordasini et al. (2008), we have considered a 25 $\mjup$ planet with a 100 $\mearth$ core. Independently of the composition of the core material (water or rock), deuterium-fusion ignition does occur in the layers
 above the core and deuterium is completely depleted in the convective H/He envelope after $\sim$ 10 Myr. The same conclusion holds for a core mass of several 100 $\mearth$.
{\it These results highlight the utter confusion provided by a definition of a planet based on the
deuterium-burning limit.}
 
\section{Planet evolutionary models and mass-radius relationships}

\subsection{The reality test}

Some of the internal structures determined in the previous sections in order to match the observed radii
and inferred mean densities of transit planets are rather unusual, and the possibility to form such compositions must be examined in the context of our current understanding of planet formation.
According to current models of planet formation, which include migration (Alibert 2005a, 2006;), up to $\sim 30\%$ of the heavy material contained in the protoplanetary disk
can be incorporated into forming giant planets (Mordasini et al. 2007).  The maximum mass
for a stable protoplanetary disk is $M_D\simle 0.1\,M_\star$, so that, for a 1 $M_\odot$ parent star of solar composition, $Z\simeq 2\%$, as much as $M_Z\approx 0.3\times 0.1\times 0.02\times (3.3\times 10^5)\approx 200\,\mearth$ of heavy material can be accreted onto the planet. Present planet formation models (Mordasini et al. 2007) reach about this limit for very massive planets, for a 1 $\msol$ parent
star. 
Therefore, {\it in principle}, according to these calculations,
a heavy element mass fraction $>50\%$ can not be excluded, even for Jovian-type planets. This
requires, however, accretion rates from the planet's feeding zone significantly larger than the values 
typical of the early "runaway growth" accretion phase, $\sim 10^{-5}\,\mearth$ yr$^{-1}$ (Ward 1996).
It also implies migration of the planet's embryo, or some substantial orbit eccentricity. Indeed, in the absence of migration or eccentricity, tidal interactions between the planet and the disk are supposed to lead
to the opening of a gap once the planet has reached about a Saturn total (gas+solid)
mass, $\sim 100 \,\mearth$, for the minimum mass solar nebulae conditions, after which planetesimal accretion decreases
dramatically (Lin \& Papaloizou 1986). Note that, if the planets formed originally at large orbital distances and migrated inwards, they are expected to have a significant content of heavy material, given the larger available
mass reservoir. Furthermore, short-period planets are expected to
have a larger fraction of heavy material than planets located further away for two reasons.
First of all, more impacts from low eccentricity orbit planetesimals are expected. Second of all,
the closer the orbit of the planet,
the larger its orbital speed, $(GM_\star/a)^{1/2}$, compared to its escape velocity, $(2GM_\star/R_\star)^{1/2}$, making planetesimal ejection less efficient (Guillot 2005). 

A second concern is the possibility to have the type of heavy element distributions examined in the present paper.
As mentioned earlier, the fact that gas accretion is triggered once the core mass has reached about 6 to
10 $\mearth$ seems to be a rather robust result (Mizuno 1980, Stevenson 1982, Pollack et al. 1996, Alibert et al. 2005a, Rafikov 2006). For larger enrichments, the rest of heavy elements should thus
be mixed with the H/He envelope. As shown by Stevenson (1982), the maximum amount of heavy material (compared with the H/He medium)
which can be redistributed uniformly by convection throughout the planet from an initially stably
stratified configuration is of the order of the planet's mass. So in principle, a planet with no or small core but
all or most of the heavy material being redistributed throughout the gaseous envelope is possible.
As shown in our study, for Jovian type planets (see \S 4.2), for a global metal fraction $Z=50$\%, if we assume a core of 10 $\mearth$, comparable to what is expected in Jupiter or Saturn, and we distribute the rest in the envelope, we find essentially the same evolution as when the heavy elements are distributed throughout the whole planet, with no core. This supposes, of course, that large-scale convection remains efficient enough to
redistribute homogeneously the heavy material in the H/He envelope (Chabrier \& Baraffe 2007). Note that this result also holds for lower mass planets  with similar enrichment,
$Z=50$\%, if the core mass remains $\simle$ 10\% $M_{\rm p}$.


Therefore, the enrichment in heavy material
and the internal compositions explored in the present calculations have at least some reasonable theoretical foundation and can not be excluded a priori.
These arguments can be examined for the case of
Hat-P-2b, for which the observed mass-radius relation requires a total mass of heavy elements of at least
200 to 300 $\mearth$, i.e. a mean mass-fraction $Z\simgr$ 7-11\%. Hat-P-2b's parent star is a F8 metal-enriched ($[Fe/H]=0.1$) star, with $M_\star=1.3\,\msol$. The maximum total amount of heavy
element material available in the parent disk was thus about 900 $\mearth$. So the required
content of heavy elements would be close to the aforementioned $\sim$30\% upper limit of accretion efficiency. 
If both the mass and the density of Hat-P-2b are confirmed, this object thus lies at the edge of what is
{\it predicted} to be possible within the current standard core-accretion scenario. We suggest, however, an alternative formation scenario, namely that the formation of Hat-P-2b
involves collision(s) with one or several other massive planets. Besides forming big cores, collisions will lead
to a substantial loss of the gaseous envelope, thus to a larger relative fraction of heavy elements. Furthermore,
gravitational scattering
among planets generally results in a tight orbit
with a large eccentricity for one of the planets, which could explain Hat-P-2b's large
eccentricity, and to the ejection of the other planet(s) or debris
to interstellar medium. Such scattering processes between planets seem to provide a viable and possibly dominant scenario
to explain the observed eccentricity distribution of exoplanets (\cite{Chatterjee07}). We thus speculate
that Hat-P-2b was formed from such collision processes. Note that a scenario based on giant impacts has also been suggested to explain the large heavy element content of HD149026b (Ikoma et al. 2006).

Finally, in the present calculations, the heavy material is supposed to be composed entirely
of one single component, water, rock (silicates) or iron. This is of course a simplifying assumption, as the inner composition of
the planets is expected to possess various fractions of each of these components. The water to silicate
fraction, in particular, will vary depending whether the object has {\it formed} inside or beyond the ice line. Migration, however, will affect this fraction, as
the migration process yields a larger collision rate of the planet embryo with rocky planetesimals and thus a decreasing
abundance of volatiles. All ratios $M_{ice}/M_{rock}$ from 0 to 1 are thus probably possible. In any event, although the exact composition of the heavy material component may have some implication on the mass-radius relationship for Earth-like planets (Valencia et al. 2006; Sotin et al. 2007; Seager et al. 2007), the
present study shows that, for planets with a gaseous H/He envelope of mass fraction $M_{HHe}\simgr$ 10\% $M_{\rm p}$, current
uncertainties on the EOS and the heavy element distribution lead to larger uncertainties on the
planet's radius determination than the effect due to different internal compositions (see Tables \ref{tab1}-\ref{tab2} and \S 5.1).

\subsection{The models}

In tables \ref{tabcond}-\ref{tabirrad}\footnote{The complete grid of models, from 10 $\mearth$ to 10 $\mjup$,  is available on http://perso.ens-lyon.fr/isabelle.baraffe/PLANET08}, we present a subset of our
grid of planetary models from 20 $\mearth$ to 1 $\mjup$ with different levels of heavy element
enrichment, $Z=Z_\odot$, 10\%, 50 \% and 90\%. For the purpose of the present paper, this grid is
 restricted for the moment to solar metallicity atmosphere models with two external atmospheric conditions, namely: (i) no irradiation (non-irradiated planets) and (ii) 
irradiation effects from a Sun at 0.045 AU, which is a typical incident irradiation for most of the transit
planets discovered up to now.
The effect of different levels of irradiation
and different atmospheric compositions will be explored in a forthcoming paper. 
We have compared our models in this mass range with the models of Fortney et al. (2007).
We find an excellent agreement for jovian-mass planets using the same assumptions, {\it i.e} same core masses and
level of irradiation. Small differences occur for Neptune-mass planets (Fortney's 17 $\mearth$ model) for a large core mass
fraction ($\mcore$ = 10$\mearth$). This is due most likely to the zero-temperature EOS assumption adopted in Fortney et al. (2007) for the core EOS, as shown earlier in this paper (see \S 4.1).
Larger discrepancies, however, appear in the irradiated case for the Neptune-mass planet.
This stems very likely from the different 
treatments of  irradiation in the atmosphere models used in Forney et al. (2007, see Fortney \& Marley 2007) and in the ones used presently (Barman et al. 2001), which seem to affect more drastically  light planets. The exploration of these effects will be considered in a future paper.

In the present grid of models, and in order to minimize the number of possibilities,  the heavy element material is restricted to {\it water} using the SESAME EOS. 
The departures from this case due to different heavy material compositions (rock, iron), distributions
within the planet's interior and EOS have been quantified in the previous sections of this paper (see Tables \ref{tab1}-\ref{tab2} for an illustration and \S 8 for a summary) for the relevant planet-mass range.
The model
users are thus referred to these analysis to determine the variations expected fom these various assumptions
in the planet modelling. Our choice for the heavy element distribution in the model grid depends on the
planet mass, and has been
guided by the most realistic expected distribution, as discussed in \S 7.1 and summarized below, from our present
analysis of the uncertainties resulting
from different treatments of heavy element enrichment. The solar metallicity models, with $Z=0.02$, have the same
composition as that of a brown dwarf, but such objects with masses as low as a few Neptune-masses
are of course not realistic. These models are an extension to planetary masses of the models of Baraffe et al. (2003).
Note that the 20 $\mearth$ planet model with an enrichment as low a $Z=0.02$ and with, consequently, a low mean density, expands
under the effect of irradiation. This model is not included in Tab. \ref{tabirrad}, being meaningless.
 \begin{itemize}
\item{} For all masses and $Z$ =  10\%, all  heavy elements are located in the core,
since for such low $Z$, we have shown that their distribution has only a modest effect
on the radius. 
\item{} For larger enrichments  ($Z$ =  50\%-90\%), we make a distinction between planets below  $M_{\rm p} \simle$ 20 $\mearth$, hereafter denominated as "light planets", and more massive planets.
Because all planets are expected
to have a core of about $\sim$ 10 $\mearth$, the following distributions seem to be the most realistic:
\begin{itemize}
\item{}  For light planets,  all heavy material is located in the core.
\item{} For more massive planets,
we have shown that distributing the heavy elements over the entire planet is similar to distributing them partly in a core of {\it at most} $\sim$ 10\% of the mass of the planet and partly in the envelope. We thus adopt such a distribution for these objects:
the heavy material is distributed over the entire planet, using the AVL with SESAME EOS. The uncertainties
due to the EOS (ANEOS versus SESAME) for this type of distribution have been quantified in \S 4 and \S 5.
\end{itemize}

\item{} For planets with masses $>$ 1 $\mjup$, we only provide models with $Z$=10\%, since enrichments 
as large as 50\% or 90\% correspond to an amount of heavy material greater than the quantity available
in protoplanetary disks around solar type stars.
 \end{itemize}
 
\begin{figure}
\psfig{file=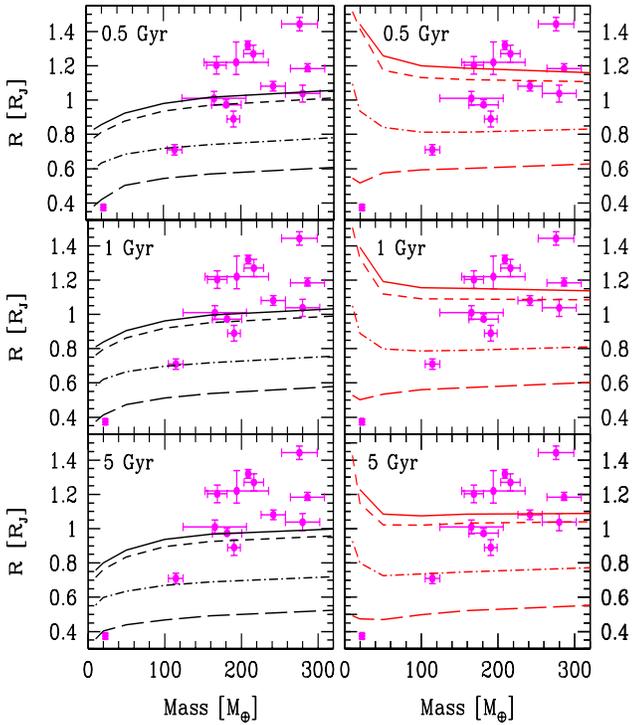,height=110mm,width=88mm} 
\caption{Mass-radius relationships for planets in the mass range 10$\mearth$ - 1 $\mjup$
at different ages, as indicated in the panels,  and different levels of heavy element enrichment: $Z=Z_\odot$ (solid lines);
$Z=$10\% (short-dash lines); $Z=$50\% (dash-dot lines); $Z=$90\% (long-dash lines). Left panels: without irradiation; right panels: with irradiation from a Sun at 0.045 AU. Symbols indicate
the observed data for transiting Neptune-mass to Jupiter mass planets, taken from F. Pont's website (obswww.unige.ch/$\sim$pont/TRANSITS.htm). }
\label{figmr}
\end{figure}

Figure \ref{figmr} shows the mass-radius relationships for planets in the mass range 20$\mearth$ - 1 $\mjup$,
for different levels of irradiation and heavy element enrichments. The models are compared to the observed mass-radius
data of transiting planets. Note that in the case of irradiation (right panel), the theoretical planetary radius is not corrected from the effect of the atmospheric extension. Such a correction
 adds $\sim$ 4\% to the calculated
radius for this level of irradiation (Baraffe et al. 2003).

\begin{table}
\caption{ Radii of planets  (in $\rjup$) in the mass range 20$\mearth$ - 1 $\mjup$ for different levels of heavy element enrichment $Z$ (see text, \S 7.2, for the choice of the heavy material
distribution) and at different ages. 
}
\label{tabcond}
\begin{tabular}{lcccc}
\hline
\hline 
$Z$ & $M_{\rm p}/\mearth$ & $R_{\rm 0.5 \, Gyr}$  & $R_{\rm 1 \, Gyr}$ & $R_{\rm 5 \, Gyr}$\\
  \hline
 \hline
0.02 & 10. &  0.828 &  0.811 &  0.758 \\
  &  20. &  0.858 &  0.839 &  0.800 \\
  &  50. &  0.923 &  0.905 &  0.876 \\
  & 100. &  0.980 &  0.963 &  0.937 \\
  & 159. &  1.017 &  0.995 &  0.968 \\
  & 318. &  1.057 &  1.032 &  0.998 \\
0.10  &  10. &  0.779 &  0.763 &  0.716 \\
  &  20. &  0.813 &  0.797 &  0.762 \\
  &  50. &  0.878 &  0.862 &  0.836 \\
  & 100. &  0.935 &  0.919 &  0.896 \\
  & 159. &  0.971 &  0.951 &  0.926 \\
  & 318. &  1.012 &  0.990 &  0.958 \\
 0.50 &  10. &  0.598 &  0.586 &  0.555 \\
  &  20. &  0.632 &  0.621 &  0.598 \\
  &  50. &  0.683 &  0.665 &  0.635 \\
  & 100. &  0.717 &  0.697 &  0.670 \\
  & 159. &  0.739 &  0.718 &  0.690 \\
  & 318. &  0.781 &  0.756 &  0.721 \\
 0.90 &  10. &  0.382 &  0.375 &  0.357 \\
  &  20. &  0.420 &  0.414 &  0.403 \\
  &  50. &  0.503 &  0.474 &  0.439 \\
  & 100. &  0.543 &  0.512 &  0.469 \\
  & 159. &  0.568 &  0.538 &  0.492 \\
  & 318. &  0.607 &  0.578 &  0.524 \\
 \hline
 \end{tabular}
\end{table}

\begin{table}
\caption{Same as table \ref{tabcond} for irradiated models by a Sun at 0.045 AU.}
\label{tabirrad}
\begin{tabular}{lcccc}
\hline
\hline 
$Z$ & $M_{\rm p}/\mearth$ & $R_{\rm 0.5 \, Gyr}$  & $R_{\rm 1 \, Gyr}$ & $R_{\rm 5 \, Gyr}$\\
  \hline
 \hline
0.02 & 10. & -  & -  & - \\
 &  20. &  1.441 &  1.391 &  1.229 \\
  &  50. &  1.258 &  1.192 &  1.084 \\
  & 100. &  1.201 &  1.155 &  1.074 \\
  & 159. &  1.186 &  1.151 &  1.085 \\
  & 318. &  1.160 &  1.137 &  1.089 \\
0.10 &  10. &  1.517 &  1.506 &  1.428 \\
  &  20. &  1.416 &  1.326 &  1.147 \\
  &  50. &  1.173 &  1.120 &  1.021 \\
  & 100. &  1.131 &  1.091 &  1.020 \\
  & 159. &  1.120 &  1.089 &  1.032 \\
  & 318. &  1.106 &  1.086 &  1.042 \\
0.50  &  10. &  1.095 &  1.050 &  0.927 \\
  &  20. &  0.935 &  0.890 &  0.800 \\
  &  50. &  0.841 &  0.798 &  0.725 \\
  & 100. &  0.812 &  0.787 &  0.737 \\
  & 159. &  0.813 &  0.790 &  0.747 \\
  & 318. &  0.830 &  0.810 &  0.772 \\
0.90  &  10. &  0.545 &  0.528 &  0.492 \\
  &  20. &  0.517 &  0.502 &  0.474 \\
  &  50. &  0.575 &  0.534 &  0.469 \\
  & 100. &  0.594 &  0.560 &  0.498 \\
  & 159. &  0.601 &  0.574 &  0.522 \\
  & 318. &  0.628 &  0.604 &  0.552 \\
 \hline
 \end{tabular}
\end{table}

\section{Discussion and perspectives}

In this paper, we have explored the uncertainties in current planetary structure and evolutionary models
arising from the treatment of the heavy material component in the {\it interior} of these planets.
The study covers
a mass range from  10 $\mearth$ to 1 $\mjup$. Our main results can be summarized as follows:
\begin{itemize}
\item{} The ideal mixing entropy contribution arising from the heavy Z-material is found to be inconsequential on the planet's {\it evolution}, within the limit that the variation of the
degree of ionization
of these elements along the evolution is presently ignored. This mixing entropy, however, represents
about 10 to 20\% of the total, H/He + Z elements, entropy, and thus modifies the internal isentrope. A
proper calculation of the planet's {\it structure} for a given entropy should thus include this contribution.
\item{} For a metal mass fraction in the envelope   $\zenv \simle$ 20\%, the EOS of the Z-material
can be approximated by using a corresponding $\yeff$ effective helium mass fraction in the
SCVH EOS. Above this limit, this approximation becomes more and more incorrect and yields
erroneous
cooling sequences.
\item{} For core mass fractions less than 50\% of the planet's mass, a variation of the core composition between
pure water and pure rock (iron) yields a difference on the radius of less than 7\% (15\%) after 1 Gyr, for all the planet masses of interest.
\item{} For a total mass fraction of heavy elements  $Z \simle$ 10\%-15\%, their impact on the
evolution of the planet can be mimicked reasonably well by assuming that they are all located in the core.
\item{} For heavy material enrichments $Z > $ 20\%, the distribution of heavy elements
(everything in the core versus uniform distribution) can affect significantly the cooling
and thus the radius determination  (more than 10\% at a given age). Therefore:
\item{} For metal-rich ($Z \simgr $ 20\%) {\it light planets} ($\simle 20 \mearth$), since the planets are expected
to have a massive $\sim 10 \mearth$ core, it seems realistic to put all the heavy material in the core.
\item{} For  {\it massive} metal-rich planets ($M_{\rm p} \simgr 50 \mearth$ and $Z > $ 20\%), however, the evolution is
better described by models which assume that the heavy elements are distributed throughout
the entire planet than by
models with all heavy elements in the core and none in the gaseous envelope. The former models yield
results similar to the ones obtained with a more realistic distribution, namely 
a  $\sim$ 10 $\mearth$  core and the rest of heavy material distributed in the envelope. 
\item{} The temperature dependence of the
heavy material EOS and the release of gravitational and thermal energy of the core have
negligible effects on the cooling history of {\it massive} planets (saturnian and jovian masses),
independently of the core mass.
For Neptune mass planets, these effects are significant ($\sim$ 10\% difference
on the radius after 1 Gyr) for extreme heavy element enrichments 
($Z > $ 90\%), as for instance in GJ 436b.
\end{itemize}

We have performed comparisons between available EOS of heavy material  which cover a range of temperature and pressure large enough to cover the characteristic domains of exoplanet interiors.
Unfortunately, significant differences exist between these EOS concerning
the thermodynamical quantities relevant for planet evolution, such as
the entropy, in particular at high pressures ($P \gg$ 0.01 Mbar) and temperatures ($T \gg $ 5000K), 
where no experimental or numerical guidance is provided. These discrepancies translate into
non-negligible variations on the mass-radius relation. 
Therefore, by venturing into parts of the phase diagram of heavy elements
 that are presently not possible to reach experimentally or
with computer numerical simulations, the planet internal structure calculations include inevitably
a certain degree of uncertainty, which has been quantified as thoroughly as possible in the present
study. For planets with even a modest fraction of H/He gaseous envelope, as expected above
$\sim 6-10\,\mearth$, these heavy material EOS related uncertainties
are the major culprit 
for preventing accurate determinations of the exoplanet internal composition from the observed mass and radius. "Accurate" means in this context at a level better than yielding $\simle 10\%$ variations on the radius.
For Earth or super-Earth planets with no gaseous envelope, a radius measurement accuracy better than 5\% is expected
to allow to distinguish icy from rocky internal compositions (Valencia et al. 2007). 
However, as shown in \S 5 for the specific cases of HD149026b and GJ436b, in the
presence of even a small ($<$ 30\% by mass) H/He contribution, varying the distribution of the heavy material
within the planet has by itself a larger impact, not mentioning the one due to the uncertainty
in the EOS.
Therefore, for planets above $\sim 10 \, \mearth$, massive enough to accrete an H/He envelope,
 it seems difficult to determine {\it precisely} the internal composition with current structure models,
as the effect of the  heavy material composition
on the radius is blurred by the presence of a gaseous envelope. 
Even a modest fraction ($\sim$ 10\%) of H/He already severely modifies the
evolution and thus the radius of a planet compared with a gasless one. As an example, a 10 $\mearth$ mass planet
retaining a 10\% H/He envelope is $\sim$ 1.5 times larger than 
its pure icy counterpart (Seager et al. 2007). 
Frustratingly enough, the
impact of this gas contribution casts doubt on our ability to determine accurately these
planet inner composition from the observed radius, at a better level than figuring out their gross bulk properties,
already an important information for our understanding of planet formation.

The main goal of this work was to explore thoroughly current uncertainties and assumptions 
in current models of planet structure and evolution. Such a study had been eluded so far in the field of exoplanets and becomes now necessary,
given the high accuracy achievable by present and forthcoming observations. 
This work enters and prolongates the long history of the modeling of giant planets, where some of the problems and 
uncertainties discussed in the present paper have been addressed decades ago for the case of our own Solar System planets. As an example,
the sensitivity of Uranus and Neptune models to the EOS of water at high pressure
has been thoroughly discussed by Hubbard \& MacFarlane (1980). The importance 
of the thermal and gravitationnal energy release of heavy 
material in planet interiors was already stressed for the evolution of
Uranus and Netpune (Podolak et al. 1991; Hubbard et al. 1995).  
All this historical work has already highlighted the remaining large uncertainties
on the modeling of our own giant planets, despite a wealth of observational constraints which are lacking  for exoplanets.

Although our conclusions
sound rather pessimistic, concerning the degree of accuracy expected from current planet models, this
work should motivate
further efforts both on the experimental and theoretical fronts, in order to make progress in this
thriving field of exoplanet exploration.
The (non exhaustive) homework list of improvements includes the following items.
(i) Exploring the EOS of heavy material
in the critical pressure regime 0.1-100 Mbar and at high temperature is crucial. Substantial progress in this domain is expected
with the ongoing and future high-pressure experiments in various national laboratories (including
e.g. the Lawrence Livermore and Sandia labs in the US or the LIL and MegaJoule laser projects in France). (ii) In the same vein, first-principle N-body numerical methods (DFT, path-integral, quantum
molecular dynamics, ...) should be able to provide at least some benchmarks in the part of the P-T
diagram which for now lies in the unknown interpolated regime. (iii) It is important to
explore the efficiency of the heat transport mechanism in planetary
interiors in the presence of molecular weight gradients, in particular in the case of large heavy material
enrichment. This bears crucial consequences on the planet's cooling history (Chabrier \& Baraffe 2007). Current progress in high-resolution multi-D numerical simulations
should be able to handle this problem.
(iv) Even though measuring the extrasolar planet gravitational moments is hardly conceivable, it might be
possible to determine their oblateness, $e=R_{eq}/(R_{eq}-R_{pol})$, where $R_{eq}$ and $R_{pol}$
denote the equatorial and polar radius, respectively, with future transit observations (Seager \& Hui 2002; Barnes \& Fortney 2003). This in turn leads to the determination of the rotation rate and thus of the centrifugal potential,
providing a more stringent constraint on the internal structure of the planet.
(v) Last but not least, the perspective, on the observational front, of
direct planetary atmosphere observations (LYOT project, GEMINI, ELT, DARWIN/TPF or their precursors)
and transit detections (CoRoT, Kepler) will improve our knowledge
 of their surface composition and radius measurements and provide important constraints on the planet's content in heavy material.

Finally, we have shown in this paper that massive ($\simgr$ several Jupiter mass) planets may form from two different avenues, namely
the standard core accretion scenario (Mordasini et al 2008), and giant impacts between massive
planets or planet embryos. This latter process is
likely to yield very metal-enriched and thus very dense massive planets, with a finite eccentricity, as a
result of planet scattering. Objects like HD 149026b, Hat-P-2b or the very recently discovered
HD 17156b (Gillon et al. 2007c) could be
the illustrations of this latter planet formation mechanism. The observation of Hat-P-2b, together with
the numerous observations of free floating brown dwarfs of a few Jupiter-masses (Caballero et al. 2007)
shows that planets and brown dwarfs have a substantial (about one order of magnitude in mass)
mass domain overlap. As we have shown in this paper, planets massive
enough to exceed the deuterium-burning mass limit will indeed ignite this reaction at the bottom of their
H/He rich envelope, at the top of the core. This is one more evidence, if it were still necessary, that
using the deuterium-burning limit as a criterion to distinguish planets  from brown dwarfs has no
valid foundation.

The complete grid of models, from 10 $\mearth$ to 10 $\mjup$,  is available on http://perso.ens-lyon.fr/isabelle.baraffe/PLANET08

\begin{acknowledgements}  
The authors thank D. Saumon for useful discussions during the elaboration of this work and our referee, J. Fortney, for his valuable comments. Part of this work was done as I.B and G.C were visiting the University of Toronto and the Max-Planck Institut for Astrophysics in Garching;  these authors thank these departments for their hospitality. The financial support of Programme National de Physique Stellaire (PNPS) and Programme National de Plan\'etologie of CNRS/INSU (France) is aknowledged.
\end{acknowledgements}

\appendix
\section{Calculation of the mixing entropy}

In this appendix, we calculate the expression for the ideal mixing entropy  
of a two-component system composed of a
H/He mixture, identified as one component, on one side and of a heavy material component of mass  
fraction $Z=M_Z/M$ on the
other side. In the following, the subscript "$1$" denotes the H/He  
component while the subscript "$2$"
refers to the Z-component. By definition, the ideal entropy of mixing  
reads:

\beq
{S_{mix}\over k_B}=\calN \ln \calN -N_1\ln N_1-N_2\ln N_2-N_e\ln N_e,
\label{ent}
\eeq
where $\calN={\cal N}_1+{\cal N}_2$ denotes the total number of  
particles, including free electrons, in the fluid, ${\cal N}_i=N_i 
+N_e^i$ denotes the total number of particles of component $i$, 
with $N_i$ the number of {\it nuclei} of component $i$ and $N_e^i$
the number of electrons provided by the component $i$. Developing eq. 
(\ref{ent}) yields:
\beq
\begin{array}{lll}
{S_{mix}\over k_B} &=& {\cal N}_1 \ln(1+{ {\cal N}_2\over {\cal N} 
_1})  +  {\cal N}_2 \ln(1+{{\cal N}_1\over {\cal N}_2})  \\
&+&  {\cal N}_1  
\ln  {\cal N}_1 +  {\cal N}_2 \ln  {\cal N}_2 \nonumber \\
& - & N_1\ln N_1-N_2\ln N_2 - N_e\ln N_e  \nonumber \\
&=& {\cal N}_1 \ln(1+{ {\cal N}_2\over {\cal N}_1}) +  {\cal N}_2 \ln(1+ 
{{\cal N}_1\over {\cal N}_2})    \\
&-& N_e\ln N_e + N_e^1\ln N_e^1 \\
& +&  N_e^2  \ln N_e^2 + {S_{mix}^{(1)}\over k_B} + {S_{mix}^{(2)}\over k_B}
\label{ent2}
\end{array}
\eeq
where

\beq
{S_{mix}^{(i)}\over k_B}={\cal N}_i \ln  {\cal N}_i -N_i\ln N_i - N_e^i\ln  
N_e^i
\eeq
denotes the ideal mixing entropy of the component $i$, including various ionic, atomic or molecular species as well
as electron contributions (see SCVH), and $N_e=N_e^1+N_e^2$ is the total number of electrons provided by the H/He and Z components. 
These ideal mixing entropy contributions are already included in the SCVH EOS for the H/He component
(their eq.(53)) 
\footnote{Note the following typos in the SCVH paper. In eqs. (45)-(46), the 
entropy ratios  on the rhs of the  
equations should be ${S^{\rm H} \over S}$ and ${S^{\rm He} \over S}$
and not ${S \over S^{\rm H} }$ and ${S \over S^{\rm He} }$.
In eq.(56) for the parameter $\delta$,  
the fraction in front of the bracketed terms on the rhs of the  
equation should be ${2\over 3}$ and not ${3\over 2}$.}
and in the appropriate EOS for the Z-component. Removing these two contributions, we obtain the ideal entropy of mixing
which arises only from the mixture of the H/He and Z components:

\beq
\begin{array}{lll}
{S_{mix}\over k_B} 
&=& {\cal N}_1 \ln(1+{ {\cal N}_2\over {\cal N}_1}) +  {\cal N}_2 \ln(1+ 
{{\cal N}_1\over {\cal N}_2})    \\
&-& N_e\ln N_e + N_e^1\ln N_e^1 
+  N_e^2  \ln N_e^2 
\label{ent2}
\end{array}
\eeq

The specific entropy, i.e. the entropy  
per unit mass is given by
\beq
{\tilde S_{mix}\over k_B}={S_{mix}\over Mk_B}= { {\cal N} \over M}
{S_{mix}\over \calN  
k_B}
\label{entspec}
\eeq
where $M=M_1+M_2$ is the total mass.

We define a mean atomic mass $\bar {m_i}=M_i/N_i$ and mean charge $1+ 
\bar {Z_i}={\cal N}_i /N_i$ for
each component, where we have used the electroneutrality condition  
$N_e^i=N_i\bar {Z_i}$,
so that ${M_i\over {\cal N}_i}={\bar {m_i} / (1+\bar {Z_i})}$. For  
the H/He component,
the quantities $\bar {m_1}$ and $\bar {Z_1}$ are given in terms of  
the relative number fractions $x$ of H$_2$, H, He, He$^+$, He$^{++}$ and e 
$^-$ by the SCVH EOS (their eqs.(33)-(35)), with:

\beq
{M_1\over {\cal N}_1}=(2x_{H_2}+x_H+x_{H^+})m_H+(x_{He^{2+}}+x_{He^+} 
+x_{He})m_{He}
\eeq
and

\beq
\bar {Z_1}={x_e^{(1)}\over 1-x_e^{(1)}}
\eeq
where

\beq
x_e^{(1)}={N_e^1\over {\mathcal N}_1}={1\over 1+\beta \gamma}x_{H^+} + {\beta \gamma \over 1+\beta \gamma}(x_{He^+}+2\,x_{He^{2+}})
\eeq
denotes the number-concentration of free electrons in the HHe mixture as defined in SCVH,
$m_{H}$ and $m_{He}$ denote the atomic mass of hydrogen and helium, respectively,
and the grec symbols $\alpha, \beta, \gamma$
are defined by eqs.(54)-(56) of SCVH. For the $Z$-component, the mean mass  
corresponds to the atomic mass
of the compound under consideration.  The mean charge $\bar Z_2$  
is unknown. We have carried out calculations for the two limiting cases
of fully neutral ($\bar Z_2=0$) and fully ionised ($\bar Z_2=Z_{nuc}$) heavy material,
where $Z_{nuc}$ denotes the nuclear charge of the compound.

After some algebra, the ideal specific entropy of mixing of the (HHe)/Z mixture can be written:

\beq
\begin{array}{lll}
{\tilde S_{mix}\over k_B} &=& {S_{mix}\over Mk_B} \nonumber \\
&=&{X_1\over \bar {m_1}} (1+\bar {Z_1}) \times
\Bigg\{
\ln (1+ \beta^\prime \gamma^\prime) + \beta^\prime \gamma^\prime \ln (1+{1 \over \beta^\prime  
\gamma^\prime}) \\
&-& x_e^{(1)} \ln (1+\delta^\prime) - \beta^\prime \gamma^\prime \, x_e^{(2)} \ln (1+{1\over  
\delta^\prime})  \Bigg\}
\end{array}
\eeq
where $X_1=M_1/M \equiv M_{HHe}/M$ and
\beq
\beta^\prime = {\bar {m_1} \over \bar {m_2} }{Z \over 1-Z},\,\,\,\,
\gamma^\prime = {1+\bar {Z_2} \over 1+\bar {Z_1} },\,\,\,\,
\delta^\prime= {N_e^2\over N_e^1}=\beta^\prime {{\bar Z_2}\over {\bar Z_1}}= {x_e^{(2)} \over x_e^{(1)}} \beta^\prime \gamma^\prime
\eeq
with $Z=M_Z/M$ and $x_e^{(i)}=\bar {Z_i} /( 1+\bar {Z_i})$.

\end{document}